\documentclass[12st]{article}

\usepackage{epsfig}

\begin{document}

\begin{center}
{\Large {\bf Neutrino opacity in magnetized hot and dense nuclear
matter}}

\vskip 25pt

{\sf Deepak Chandra $^{\dagger}$}
\footnote{E-mail address: dchandra@himalaya.du.ac.in }, 
{\sf Ashok Goyal 
\footnote{E-mail address: agoyal@ducos.ernet.in}}
and  
{\sf Kanupriya Goswami $^{\ddag}$}  
\\ {\em Department of Physics \&  Astrophysics, University of Delhi, Delhi 110 007, India}   \\ {\em \& Inter University Centre for Astronomy \&  Astrophysics,
Ganeshkhind, Pune-411007, India}\\

$^a${\em  Department of Physics, S.G.T.B. Khalsa College, University of
Delhi, Delhi 110 007, India }\\

$^b${\em Department of Physics, Keshav Mahavidyalaya, Keshavpuram, 
Delhi 110035, India}\\

\end{center}
\vskip 20pt 

\begin{abstract}
\noindent We study the neutrino interaction rates in hot matter at
high densities in the presence of uniform magnetic field. The neutrino
cross-sections involving both the charged current absorption and
neutral current scattering reactions on baryons and leptons have been
considered.  We have in particular considered the interesting case
when the magnetic field is strong enough to completely polarise the
protons and electrons in supernovae and neutron stars. The opacity in
such a situation is considerably modified and the cross-section
develops anisotropy. This has implications for phenomenon invoked in
the literature to explain the observed pulsar kicks.
\end{abstract}

\vskip 1 true cm
\pagebreak

\begin{section} {\bf Introduction}
\indent Knowledge of neutrino transport in supernovae cores, in
neutron stars and in collapsing stars is an essential prerequisite for
an understanding of a host of interesting phenomenon like the
mechanism of supernovae explosion, structure of proto-neutron stars,
observed pulsar kicks etc. The important theoretical input required
for the study of neutrino transport is neutrino opacity calculations
in dense, hot matter and the equation of state (EOS). There has been a
lot of work on neutrino interactions in matter at high densities
{\cite{Red}} involving both the charge current absorption and neutral
current scattering reactions on baryons and leptons. The dominant
processes for energy and lepton number transport are the
neutrino-nucleon scattering and the neutrino absorption. It has been
shown that neutral current processes on charged leptons, though
subdominant with regard to total opacity are important for the
equilibrium of the neutrino number density.  \par Recently there has
been a lot of interest in the study of neutrino transport at high
densities in the presence of strong magnetic fields reported to be
present in young pulsars. Presence of strong magnetic fields of
strengths $\sim 10^{14}$ Gauss have been suggested in neutron stars
and recent observations of soft gamma ray repeaters and spinning X'ray
pulsars even suggest the existence of magnetic fields greater than
$10^{14}$ Gauss in supernovae remnants {\cite{Kouv}}. A study of
neutrino opacities in magnetised high densitiy, hot matter is required
to investigate the asymmetric neutrino emission from proto-neutron
stars as a possible explanation of observed pulsar kicks
{\cite{Lyne}}.  Among various mechanisms suggested for pulsar kicks,
the notable ones relevant to the present study are the models in which
the magnetic field induces an assymetry directly in the neutrino
emission {\cite{Vilenk}} giving rise to the observed kicks and those
which rely on resonant M-S-W flavor transformation on account of
changed resonance condition in the magnetic field {\cite{Kusen}} or on
neutrino magnetic moment {\cite{Akhve}}. This requires a careful,
systematic study of neutrino interactions in hot, dense, magnetised
matter in various stages of degeneracy. For example, electron
-neutrino absorption and scattering cross-sections on nucleons and
leptons would not only shift the location of the $\nu_e$ sphere but
will also distort it due to asymmetry in the cross-section. This
effect on the dominant reaction $\nu_e+n\rightarrow p+e$ at densities
near the electron-neutrino sphere was calculated by multyplying the
phase space distribution of the final state electrons only and leaving
the matrix elements unchanged {\cite{Esteb}}. However it was noted
{\cite{Ashok}} that in the presence of large magnetic field $\sim
10^{16}$ Gauss even the motion of degenerate electrons, at densities
likely to be present near the neutrino sphere, is quantised and on
account of charge neutrality viz. $n_e=n_p$, the protons too are
forced to occupy the lowest Landau level. In this situation, the
matrix elements for absorption and scattering get modified and have to
be calculated by using exact wave functions for electrons and protons
by solving the Landau level $\nu =0$ {\cite{Ashok}}. Further, in order
to make numerical estimates of neutrino mean free paths in magnetised
dense and hot nuclear matter in the density range $\sim 10^{10} -
10^{15}$ g/cc and temperatures upto $60$ MeV, we require the
composition of nuclear matter. The effect of magnetic field on the
composition of nuclear matter at low densities has been extensively
studied in the literature {\cite{Fushiki}}. In recent years the effect of
strong magnetic field on cold, charge neutral, superdense interacting
nuclear matter in $\beta$-equilibrium has been studied in a
relativistic mean field theoretical frame work [10-12]. In some of
these studies [11,12] not only the effect of Landau quantisation but
also the contribution of anamalous magnetic moments of nucleons was
incorporated in a relativistic description and it was found that this
effect cannot be ignored at low densities in the presence of
superstrong magnetic field $\geq 10^{18}$ Gauss.  Here following
reference {\cite{Agoyal}} we consider electrically neutral nuclear matter
composed of nucleons, electrons and trapped neutrinos in $\beta $-
equilibrium. Hadronic interactions are incorporated through ($\rho -
\omega - \sigma $ ) meson exchange in the framework of relativistic
nuclear mean field theory in the presence of magnetic field. When
considering very intense magnetic fields we should be careful about
the effect of magnetic field on strong interactions because
in this situation, the nuleons and mesons interact both with
themselves, and with the magnetic field through their charges and
magnetic moments. However, for fields not greater than $10^{18}$ gauss
we do not have to worry about this problem.  \par In this paper, we
calculate neutrino opacity for magnetised, interacting dense nuclear
matter for the following limiting cases: a) nucleons and electrons,
highly degenerate with or without trapped neutrinos, b) non-degenerate
nucleons, degenerate electrons and no trapped neutrinos and finally,
c) when all particles are non-degenerate. The important neutrino
interaction processes which contribute to opacity are the neutrino
absorption process
\begin{equation}
\nu_e + n\rightarrow p + e
\end{equation}
and the scattering processes
\begin{equation}
\nu_e + N\rightarrow \nu_e + N
\end{equation}
\begin{equation}
\nu_e + e\rightarrow \nu_e + e
\end{equation}
\par In section II we calculate the cross-sections for these processes
in the presence of magnetic field. In section III the cross-sections
are calculated in the polarised medium when the magnetic field is very
strong and density is such that the electrons and protons are confined
to the lowest Landau level. For the purpose of calculating reaction
rates we treat the nucleons non-relativistically and the leptons in
extreme relativistic limit. In section IV we discuss
the results.
\end{section}

\vskip 1.0 cm
\begin{section}{\bf Neutrino Cross-section}
\noindent The neutrino processes (1-3) get contribution from charged
as well as neutral current weak interactions in the standard
model. For the general process

\begin{equation}
\nu (p_1) + A(p_2)\rightarrow B(p_3) + l(p_4)
\end{equation}

The cross-section per unit volume of matter or the inverse mean free
path is given by

\begin{equation}
\frac {\sigma (E_1)} {V} = \lambda ^{-1}(E_1) = \frac {1} {2E_1} \prod
_{i=2,3,4} d\rho_i W_{fi} f_2(E_2)(1-f_3(E_3))\left(1-f_4(E_4)\right)
\end{equation}

where $d\rho _i=\frac {d^3p_i} {(2\pi )^32E_i}$ is the density of
states of particles with four momenta $p_i=(E_i,\vec{p}_i)$,
$f_i(E_i)$ are the particle distribution functions which in thermal
equilibrium are given by the usual Fermi-Dirac distributions
$f_i(E_i)=[1+e^{ \beta (E_i-\mu _i)}]^{-1}$ and $1-f_i(E_i)$ accounts
for the Pauli-Blocking factor for the final state particles. $\mu
_i$'s are the chemical potentials and $\beta =1/(KT)$.  The transition
rate $W_{fi}$ is

\begin{equation}
W_{fi} = (2\pi )^{4} \delta ^{4} (P_f-P_i)\left |M\right |^2
\end{equation}

where $\left |M\right |^2$ is the squared matrix element summed over
the initial and final spins.

\par

In the presence of magnetic field, the density of states of charged
particles is modified and is given by

\begin{equation}
d\rho _i = \sum _{\nu } (2-\delta _{\nu ,0})\int_{-\infty}^{\infty}
dp_{iz}\int_{-eBL_x/2} ^{eBL_x/2} \frac {dp_{iy}} {(2\pi )^2 2E_i}
\end{equation}

where the sum over all occupied Landau levels is to be performed.  For
weak magnetic fields, several Landau levels are populated and the
matrix element remain essentially unchanged {\cite{Bezchastnov}} and one
needs to only account for the correct phase space factor. In the
presence of strong magnetic field, the electrons occupy the lowest
Landau state and the charge neutrality requirement then forces the
protons into the ground state. This statement hold for cold degenerate
matter. For hot matter with temperatures $\geq 10$ MeV and density
$10^{11} - 10^{12}$ g/cc, the protons are not so severly affected by
the magnetic field. This happens because of the presence of a sizable
number of positrons at these densities and temperatures, thereby
modifying the charge neutrality condition.  In the presence of such
intense magnetic field, the matrix elements have to be calculated by
using the exact wave functions of relativistic electrons and protons
obtained by solving the Dirac equation.  \par In a gauge in which the
vector potential is $\vec{A}=(0,xB,0)$, corresponding to a constant,
uniform magnetic field in the z-direction, the quantum states are
specified by the quantum numbers $p_y$, $p_z$, $\nu$ and $S$ and the
energy eigenvalues is given by [11,12]
\begin{equation}
{\cal E}_{\nu ,s}^{e} = {\left[m^2 + p^2_z + eB(2 n + 1 +
s)\right]}^{\frac {1} {2}}
\end{equation}

\begin{eqnarray}
{\cal E}_{\nu,s}^{p} &=& \sqrt{m_{p}^{*2}+p_{z}^2+eB(2\nu+1+s)
+\kappa_{p}^2 B^2+2\kappa_{p} B
s\sqrt{m_{p}^{*2}+eB(2\nu+1+s)}}\nonumber \\ &=&
E_{\nu,s}^{p}-U_{0}^{p}
\end{eqnarray}

and
\begin{eqnarray}
{\cal E}_{s}^{n} &=&
\sqrt{m_{n}^{*2}+\overrightarrow{p^{2}}+\kappa_{n}^2 B^2 + 2\kappa_{n}
B s\sqrt{p_{x}^{2}+p_{y}^{2}+m_{n}^{*2}}} \nonumber \\ &{}& =
E_{s}^{n}-U_{0}^{n}
\end{eqnarray}
where $\kappa_p$ and $\kappa_n$ are the anamalous magnetic moments of
protons and neutrons being given by

\begin{equation}
\kappa_{p}=\frac{e}{2m_{p}}[\frac{g_{p}}{2}-1]\nonumber \\
\end{equation}

\begin{equation}		
\kappa_{n}=\frac{e}{2m_{n}}\frac{g_{n}}{2}\nonumber \\
\end{equation}

where $g_{p}=5.58$ and $g_{n}=-3.82$ are the Lande's g-factor for
protons and neutrons respectively. We have here neglected the
contribution of anamalous magnetic moment of electrons. $U_{0}^{p}$,
$U_{0}^{n}$, $m_{N}^\star$ are given in the appendix A.

$\nu =n+\frac{1}{2}\pm \frac{s}{2}$ characterises the Landau level.
The electron wave function in the lowest Landau state $\nu=0$ has the
energy $E_{0,-1}^{e}=(m_{e}^2 +p_{z}^{2})^\frac{1}{2}$ with the wave
function
\begin{equation}
\Psi (r) = \left (\frac {eB} {\pi}\right )^{\frac {1} {4}} \frac {1}
{\sqrt {L_yL_z}}e^{-i(E_{e}t - p_yy - p_zz)} e^{\xi ^2/2}
U_{e,1}(E_{e})
\end{equation}
and the positve energy spinors in state $\nu $=0 are given by


\begin{equation}
U_{e,-1} = \frac {1} {\sqrt {E_e + m_e}}\pmatrix {0 \cr E_e+m_e \cr 0
\cr -p_z}
\end{equation}
Nucleons are treated non-relativistically. In these limits the proton
energy in the lowest state is given by
\begin{eqnarray}
E_{0,1}^{p} &=& \frac {m_{p}^\star-\kappa_{p}B}{\bar m_{p}} +
\frac{p_{z}^{2}}{2(m_{p}^\star-\kappa_{p}B)}+U_{0}^{p}\nonumber\\
\end{eqnarray}
where
\begin{equation}
\bar m_{p}=m_{p}^\star-\kappa_{p}B
\end{equation}
and the wave function is
\begin{equation}
\Psi_{p} (r) = \left (\frac {eB} {\pi}\right )^{\frac {1} {4}} \frac
	{1} {\sqrt {L_yL_z}}e^{-i(E_{0 ,1}^{p}t - p_yy - p_zz)} e^{\xi
	^2/2} U_{p,1} ( E_{0,1}^{p})
\end{equation}
where $U_{p,1}( E_{0,1}^{p})$ is the non-relativistic spin up
spinor.
 
\begin{equation}
U_{p,1}( E_{0,1}^{p}) = \frac {1} {\sqrt {2m_p}}\pmatrix {\chi_+
\cr 0 \cr}
\end{equation}
and
\begin{equation}
\xi =\sqrt{eB} (x-\frac{p_{y}}{eB})
\end{equation}

For neutrons we have
\begin{equation}
\Psi_{n,s} (r) = \frac {1} {\sqrt {L_xL_yL_z}}
	e^{-ip_{n,s}(r)}U_{n,s}( E_{s}^{n})
\end{equation}

\begin{equation}
U_{n,s} = \frac {1} {\sqrt {2m_n}}\pmatrix {\chi_s \cr 0 \cr}
\end{equation}

and
\begin{equation}
E_{n,s} \simeq \frac{m_{n}^{\star}-\kappa_{n}Bs}{\bar m_{n}}
	+\overrightarrow{p^{2}}+U_{0}^{n}
\end{equation}

with
\begin{equation}
\bar m_{n}=m_{n}^{\star}-\kappa_{n}Bs
\end{equation}
Neutrino wave function is the usual plane wave

\begin{equation}
\Psi_{\nu} (r) = \frac {1} {\sqrt {L_xL_yL_z}}
	e^{-ip_{\nu}(r)}U_{\nu,s}(E_{\nu})
\end{equation}

Here $U_{\nu,s}$ is the usual free particle spinor.  $\chi_{s}$ is the
spin spinor and the wave functions have been normalised in a volume
$V=L_xL_yL_z$ and we have the normalization

\begin{equation}
\sum _{Spin} \bar{u}_{\alpha}u_{\beta} = 2m\delta _{\alpha \beta}
\end{equation}

\par We first consider the neutrino-nucleon processes (1,2). In the
presence of weak magnetic fields, the matrix element squared summed
over initial and final spins in the approximation of treating nucleons
non-relativistically and leptons relativistically is given by
\begin{equation}
\sum \left |M\right |^2 =
32G_{F}^{2}\cos_{\theta_{c}}^{2}m_p^{\star}m_n^{\star}\left[(C_{V}^{2}) + 3C_{A}^{2})
+ (C_V^2 - C_A^2)cos\theta _{e,\nu }\right]E_eE_{\nu }
\end{equation}
where $C_V=$ $g_V=$1, $C_A=$ $g_{A}=$1.23 for the absorption process;
      $C_V=-$1, $C_A=-$1.23 for neutrino scattering on neutrons and
      $C_V=-1+ 4sin^{2}\theta _{w}=$0.08, $C_A=$1.23 for netrino
      proton scattering.  \par
\noindent The effect of strong interactions on charged and neutral
current neutrino interactions is incorporated in the framework of
relativistic mean field theory. The nuclear matter consisting of
neutrons, protons, electrons with or without neutrinos is considered
in $\beta$- equilibrium in the presence of magnetic field. Hadronic
interactions are taken into account by considering the nucleons to
interact by exchanging scalar $(\sigma)$ and vector $(\omega , \rho)$
mesons {\cite{Agoyal}}.  The composition of matter is calculated for
arbitrary magnetic fields and temperatures without making any
approximation regarding relativity and degree of degeneracy of various
particles. For the effect on weak rates, we however treat the nucleons
non-relativistically and obtain neutrino cross-sections in the limits
of extreme degeneracy or for non-degenerate matter. This simplifies
the calculations considerably without compromising the results in any
significant way. This results in replacing the nucleon mass $m$
by it's density dependent mass $m^*$ and the chemical potential by the
appropriate effective chemical potential (see Appendix A).

\noindent We present here the results for interacting nuclear matter
consisting of neutrons, protons, electrons and neutrinos(for the trapped case
where lepton fraction $Y_{L}=Y_{e}+Y_{\nu_{e}}$ is held fixed at 0.4) for
nuclear density $n_{B}$ varying over a wide range($10^{-5}-10$)$n_{0}$ 
and at temperatures T=5, 10 and 30 MeV. In figures 1 and 2 we show the effect
of magnetic field (measured in the units of $B_{e}^{c}$ ,viz., 
$B^{\star}=\frac{B}{B_{e}^{c}}$ and the critical magnetic field for 
electrons is$B_{e}^{c}=4.413 \times 10^{13}$ G) on the composition of 
nuclear matter viz 
$Y_{p}$, $Y_{n}$ and $Y_{\nu_e}$ as a function of nuclear density
for T=5, 30 and 60 MeV 
for the neutrino trapped and untrapped cases respectively.
The effect of magnetic field is to raise the proton fraction and is very
pronounced at low densities. At densities relevant to the core of neutron 
stars, one requires very high magnetic fields to change the 
composition. The effect of including anamalous magnetic moment also becomes 
significant at field strengths typically $\geq 10^{5}$ which can be seen 
in Fig 3 where we have plotted the proton and neutron fractions 
$Y_{p}$ and $Y_{n}$ and the  effective nucleon mass as a 
function of density for large magnetic field $5\times 10^5$ with and without 
including the effect of anomalous magnetic moment of nucleons.

\begin{subsection}{Highly degenerate matter}
\noindent The absorption cross-section (1) for this case can be
calculated by using (14) in (5) by the usual techniques and we get for
small B:-
$$\frac {\sigma_{A} (E_{\nu },B)} {V} = \frac {G^2_F
\cos_{\theta_{c}}^{2}} {8\pi ^3}(g^2_V + 3g^2_A)m^*_pm^*_nT^2 \frac
{\left(\pi^2 + \frac {(E_\nu - \mu_{\nu})^2} {T^2}\right)} {\left(1 +
e^{ \frac {(\mu_{\nu } - E_{\nu} )} {T}}\right)} eB[\theta (p_F(p) +
p_F(e)$$
$$- p_F(n) - p_F(\nu )) + \frac {(p_F(p) + p_F(e) - p_F(n) + p_F(\nu
)} {2E_\nu }(\theta (p_F(\nu )$$
\begin{equation} 
- \left |p_F(p) + p_F(e) - p_F(n)\right |)]\sum _{\nu =0}^{\nu
  _{max}}(2-\delta _{\nu,0})\frac {1} {\sqrt {\mu^2_e - m^2_e -2\nu
  eB}}
\end{equation}
which reduces to the usual result [13] in the limit $B\rightarrow 0$.
The case of freely streaming, untrapped neutrinos is obtained from
(27) by putting $\mu_{\nu }=$0 and replacing $\mu_{e}$ by
$(\mu_{e}+E_{\nu})$.  

\par

When the magnetic field is much weaker than the critical field for
protons, only electrons are affected and the neutrino-nucleon
scattering cross-section expression remain unchanged by the magnetic field. 
The numerical values 
however, are modified due to changed chemical composition. 
The cross-sections are given by

\begin{equation}
\frac {\sigma _{\nu N}(E_{\nu})} {V} = \frac {G^2_F\cos_{\theta_{c}}^{2}} {16\pi ^3}(C^2_V + 3C^2_A){m^*_N}^2T^2 \mu_{e} \frac {\left(\pi^2 + \frac {(E_\nu - \mu _\nu)^2} {T^2}\right)} {\left(1 + exp\frac {(E_\nu - \mu _\nu)} {T}\right)}
\end{equation}

If neutrinos are not trapped, we get in the elastic limit

\begin{equation}
\frac {\sigma _{\nu N}(E_\nu)} {V} = \frac {G^2_F\cos_{\theta_{c}}^{2}} {16\pi ^3}(C^2_V + 3C^2_A){m^*_N}^2T^{2}E_\nu
\end{equation}

The neutrino electron scattering though important for energy momentum 
transfer, is subdominant for neutrino opacity calculations which get 
major contribution from the absorption reaction in the neutrino trapped 
regime and by absorption and scattering processes on nucleons in the other 
regime. The electron-neutrino scattering in the presence of arbitrary 
magnetic field has been evaluated in the literature by [14]. Here we 
will consider the interesting case when the electrons get totally polarised 
by the magnetic field. For completeness, we give below the 
neutrino-electron scattering cross-section in the absence of the field 
for relativistic degenerate electrons and neutrinos in the elastic 
approximation.
\begin{equation}
\frac {\sigma _{\nu e}} {V} \simeq \frac {2G_F^2\cos_{\theta_{c}}^{2}} {3\pi ^3}(C_V^2 + C_A^2)\frac {\mu _e^2 TE_\nu ^2}{1 + e^{-\beta (E_\nu - \mu_\nu )}}
\end{equation}
which in the untrapped regime goes over to
\begin{equation}
\frac {\sigma _{\nu e}} {V} \simeq \frac {2G_F^2\cos_{\theta_{c}}^{2}} {3\pi ^3}(C_V^2 + C_A^2)\mu _e^2 TE_\nu ^2
\end{equation}
\end{subsection}
\begin{subsection}{Non-degenerate matter}
\noindent We now treat the nucleons to be non-relativistic non-degenerate 
such that $\mu_i/T \ll -1$ and thus the Pauli-blocking factor $1-f_N(E_i)$ 
can be replaced by 1, the electrons are still considered degenerete and 
relativistic. In the approximation which is valid near the neutrino spheres 
for the proto-neutron star in the Helmholtz-Kelvin cooling phase, 
the cross-section

$$\frac {\sigma_{A} (E_\nu ,B)} {V} \simeq \frac {G_F^2\cos_{\theta_{c}}^{2}} {2\pi }(C_V^2 + 3C_A^2)n_n (E_\nu+Q)  \frac {1}{1 + e^{-\beta (E_\nu +Q - \mu_e )}}$$
\begin{equation}
eB\sum _{\nu =0}^{\nu _{max}}(2-\delta _{\nu,0})\frac {1} {\sqrt{(E_{\nu}+Q)^{2} - m_e^2 -2\nu eB}}
\end{equation}
where $n_N$ is the nuclear desity and $Q=m_n - m_p$.
\begin{equation}
\frac {\sigma_{\nu N} (E_\nu ,B)} {V} \simeq \frac {G_F^2\cos_{\theta_{c}}^{2}} {4\pi }(C_V^2 + 3C_A^2)n_NE_\nu ^2
\end{equation}

If the electrons too are considered non-degenerate we get
\begin{equation}
\frac {\sigma _{A}} {V} \simeq \frac {G_F^2\cos_{\theta_{c}}^{2}} {\pi }(g_V^2 + 3g_A^2)n_NE_\nu ^2
\end{equation}

\begin{equation}
\frac {\sigma _{\nu N}} {V} \simeq \frac {G_F^2\cos_{\theta_{c}}^{2}} {4\pi }(C_V^2 + 3C_A^2)n_NE_\nu ^2
\end{equation}

\begin{equation}
\frac {\sigma _{\nu e}} {N} \simeq \frac {4G_F^2\cos_{\theta_{c}}^{2}} {\pi ^3}T^4E_\nu (C_V + C_A)^2
\end{equation}

\end{subsection}
\end{section}

\begin{section}{\bf Neutrino cross-sections in polarised medium}

\noindent When the magnetic field exceeds $p_F^2(e)/2$, all the electons 
in matter occupy the lowest Landau state $\nu =$0 with their spins 
pointing in the direction opposite to the magnetic field. In this 
situation charge neutrality for degenerate matter forces the 
non-relativistic protons to be also in the Landau ground state but with 
their spins alinged along the magnetic field. In this situation matrix 
elements cannot be considered unchanged and should be evaluated using the 
exact solutions of Dirac equation for charged particles in magnetic field. 
Using the wave functions given in section 2, the square of the matrix elements for weak 
processes can be evaluated in a straight forward way and we get
$$\left |M\right |^2_{A} = 8G_F^2cos^2\theta _cm^*_2m^*_3(E_4+p_{4z})\left[(g_V + g_A)^2(E_1 + p_{1z}) + 4g^2_A(E_1 - p_{1z})\right]$$
\begin{equation}
exp\left[ - \frac {1} {2eB}((p_{1x} + p_{2x})^2 + (p_{3x} + p_{4y})^2)\right]
\end{equation}
\begin{equation}
\left |M\right |^2_{\nu p } = 16G^2_F\cos_{\theta_{c}}^{2}C^2_A{m^*_2}^2(p_1\cdot p_4 + 2p_{1z}p_{4z})exp\left[-\frac {1} {2eB}((p_{4x} + p_{1x})^2 + (p_{4y} - p_{1y})^2)\right]
\end{equation}
$$\left |M\right |^2_{\nu e } = 16G^2_F\cos_{\theta_{c}}^{2}[(C^2_V + C^2_A)((E_1E_4 + p_{1z}p_{4z})(E_2E_3 + p_{2z}p_{3z})$$
$$- (E_1p_{4z} + E_4p_{1z})(E_2p_{3z} + E_3p_{2z}))$$
$$+ 2C_VC_A((E_1E_4 + p_{1z}p_{4z})(E_2p_{3z} + E_3p_{2z}) - E_1p_{4z} + E_4p_{1z})$$
\begin{equation}
(E_2E_3 + p_{2z}p_{3z}))]exp\left(- \frac {1} {2eB}((p_{4x} - p_{1x})^2 - (p_{4y} - p_{1y})^2)\right)
\end{equation}
The absorption cross-section is now given by
$$\frac {\sigma _{A}(E_1,B)} {V} = \frac {1}{2E_1L_x} \int \frac {d^3p_z} {(2\pi )^32E_2}\int_{-eBL_x/2} ^{eBL_x/2}\int_{-\infty }^{\infty } \frac {dp_{3y}dp_{3z}} {(2\pi )^22E_3}$$
$$\int_{-eBL_x/2} ^{eBL_x/2}\int_{-\infty }^{\infty } \frac {dp_{4y}dp_{4z}} {(2\pi )^22E_4}(2\pi )^3\delta (P_y)\delta (P_z)\delta (E)\left |M\right |^2$$
\begin{equation}
f_2(E_2)(1-f_3(E_3))(1-f_4(E_4))
\end{equation}
The corresponding scattering cross-sections are obtained by interchanging 
the particle 2 with 4. Integrals over $dp_{3y}dp_{4y}$ can be performed 
by using the $y$-component momentum conserving $\delta (P_y)$ function and 
give $eBL_x$. $d\Omega _2$ integral can be performed by using 
$\delta (P_z)=\frac {1} {p_2}\delta \left(cos\theta _2-\frac {p_{3z}+p_{4z}-p_{1z}} {p_2}\right)$ and gives
$2\pi $. Here for simplification we have replaced the angles appearing in 
the exponential factor by their average values. It is not a bad approximation 
since the exponential factor is indeed of order one in the high field limit. 
We thus obtain
$$\frac {\sigma _{A}(E_1,B)} {V} \simeq  \frac {eB} {2E_1} \frac {1} {(2\pi )^3} \frac {1} {8}\int dE_2\frac {dp_{3z}} {E_3}\frac {dp_{4z}} {E_4}\delta (E)\left |M\right |^2$$
\begin{equation}
f_2(E_2)(1-f_3(E_3))(1-f_4(E_4))
\end{equation}
To make further progress, we consider the cases of extreme-degeneracy 
and non-degeneracy separately.
\vskip 1.0 cm 
\begin{subsection}{Highly Degenerate matter}
\noindent We first convert the integrals over $z$-component of the 
electron and proton momentum into integrals over the electron and proton 
energy respectively by using the energy expression (8) for $\nu =$0. 
Further, since for strongly degenerate matter, particles at the top of 
their respective fermi-seas alone contribute, we replace the momenta by 
their fermi momenta. In this approximation we get
$$\frac {\sigma _{A}(E_1,B)} {V} \simeq  \frac {G^2_Fcos^2\theta _c} {(2\pi )^3}eB\frac {m^*_3m^*_4} {p_F(3)E_1}\frac {T^2} {2}\frac {\left(\pi ^2 + \left(\frac {E_1} {T}\right)^2\right)} {(1 + e^{-E_\nu /T})}\int dE_2dE_3dE_4$$
$$\delta (E_1 + E_2 - E_3 - E_4)f_2(1-f_3)(1-f_4)\left[(g_V + g_A)^2(E_1 + p_{1z}) + 4g^2_A(E_1 - p_{1z})\right]$$
$$[exp\left(-\frac {\left(p_F(1) + p_F(2)\right)^2 - \left(p_F(3) - p_F(4)\right)^2} {2eB}\right)$$
$$\theta \left((p_F(1) + p_F(2))^2 - (p_F(3) - p_F(4))^2\right)$$
$$+exp\left(-\frac {(p_F(1) + p_F(2))^2 - (p_F(3) + p_F(4))^2} {2eB}\right)$$
\begin{equation}
\theta\left((p_F(1) + p_F(2))^2 - (p_F(3) + p_F(4))^2\right)]
\end{equation}

Carrying out the energy integrals 
\begin{eqnarray}
\frac {\sigma _{A}(E_{\nu},B)} {V} 
&\simeq &  \frac {G^2_Fcos^2\theta _c} {(2\pi )^3}eB\frac {m^*_nm^*_p} 
	{p_F(\nu )E_\nu }\frac {T^2} {2}\frac {(\pi ^2 
	+ (\frac {E_\nu } {T})^2)} 
	{(1 + e^{-E_\nu /T})}\nonumber \\
&&{}	[(g_V + g_A)^2(E_\nu  + p_{\nu z}) 
	+ 4g^2_A(E_\nu  - p_{\nu z})]\nonumber \\
&&{}	[exp(-\frac {(p_F(n) + p_F(\nu ) )^2} {2eB} ) 
	\theta (p_F(n) + p_F(p ) )\nonumber \\
&&{}	+exp(-\frac {(p_F(n) + p_F(\nu ))^2 - 4p^2_F(e)} {2eB})\theta ((p_F(n) + p_F(\nu ))^2 - 4p^2_F(e))]\nonumber \\
\end{eqnarray}
where we have made use of the fact that for electrically neutral matter 
$p_F(e)=p_F(p)$ and that two $\theta $ functions correspond to the fact 
that $z$-component momentum conservation 
$p_{nz}+p_{\nu z}=p_{ez}+p_{pz}=p_F(e)\pm p_F(p)$ depending on whether 
both electrons and protons in their Landau ground state  in the same 
direction or opposite to each other. The case of freely streaming 
neutrons is obtained from the above by putting $p_F(\nu )=$0 everywhere.
\par
For the case of neutrino-proton scattering, we can perform the integrals 
in the degenerate limit by using 
$\int _{}^{}dp_{2z}$ as $\int _{}^{}dp_{2z}d\delta 
\left( \frac {p^2_{2z}} {2m^*_2}-\frac {p^2_f(2)} {2m^*_2}\right)d\epsilon_{2}$ 
and likewise for 
$\int _{}^{}dp_{2z}$, where $\epsilon_{i}=\frac{p_{iz}^{2}}{2 m_{i}^{\star}}$. $d^3p_4$ integral can be performed by choosing $\vec{p_1}=E_1(0,sin\theta _1,cos\theta _1)$ and $\vec{p_4}=E_4(sin\theta _4cos\phi _4,sin\theta _4sin\phi _4,cos\theta _4)$. $d(cos\theta _4)$ integral is performed as before by using $\delta (P_z)$ and noticing that $cos\theta _4=\frac {p_{1z}} {p_4}$ alone is allowed when initial and final protons move along the same direction, in the oposite case $cos\theta _4=\frac {2p_F(p)+p_{1z}} {p_{4z}} >$ 1 and thus not allowed. Approximating the exponential factor by 1 we get
$$\frac {\sigma _{\nu p}(E_1,B)} {V} \simeq \frac {G^2_Fcos^2\theta_c} {(2\pi )^3}\frac {{m^*_2}^2} {p^2_F(2)}\frac {eB} {2}\int d\epsilon_2d\epsilon_3dE_4\left(E_4 + \frac {p^2_{1z}} {E_1}\right)$$
\begin{equation}
f_2(1-f_3)(1-f_4)\delta (E_1 + \epsilon_2 -\epsilon_3 - E_4)
\end{equation}
$$\simeq \frac {G^2_Fcos^2\theta _c} {(2\pi )^3}\frac {{m^*_p}^2} {p^2_F(p)}g^2_A\frac {eB} {2}\frac {T^2} {2}$$
\begin{equation}
\frac {(\pi ^2 + (\frac {E_\nu - \mu_\nu} {T})^2)} {(1 + e^{-(E_\nu - \mu _\nu)/T})}\left(\mu _\nu + \frac {p^2_{\nu z}} {E_\nu}\right)
\end{equation}
\begin{equation}
\simeq \frac {G^2_Fcos^2\theta _c} {64\pi }\frac {{m^*_p}^2} {p^2_F(p)}g^2_AeBT^2\mu _\nu(1 + cos^2\theta)
\end{equation}
because $E_\nu\simeq \mu _\nu$ and $\theta$ is the angle which 
neutrino makes with the magnetic field. For the case when neutrinos 
are not trapped we get
\begin{eqnarray}
\frac {\sigma _{\nu p}(E_\nu ,B)} {V} 
& \simeq &\frac {G^2_Fcos^2\theta _c} {(2\pi )^3}
	\frac {{m^*_p}^2} {p^2_F(p)}g^2_A\frac {eB} {2} \int 
	\left(E_4 + \frac {p^2_{1z}} {E_1}\right)(E_1 - E_4)
	\frac {dE_4} {1 - e^(E_1 - E_4)/T}\nonumber \\
\end{eqnarray}
\begin{equation}
\simeq \frac {G^2_Fcos^2\theta _c} {(2\pi )^3}\frac {{m^*_p}^2} {p^2_F(p)}g^2_AT\frac {E^2_\nu} {2}\frac {eB} {2}(1 + 2cos^2\theta )
\end{equation}
Likewise for the neutrino-electron scattering we get
\begin{equation}
\frac {\sigma _{\nu e}(E_\nu ,B)} {V} \simeq \frac {G^2_Fcos^2\theta _c} {8\pi }T^2eB\mu _\nu \left[(C^2_V + C^2_A)(1 + cos^2\theta ) - 4C_VC_Acos\theta \right]
\end{equation}
for the trapped neutrino case and
\begin{equation}
\frac {\sigma _{\nu e}(E_\nu ,B)} {V} \simeq \frac {2G^2_Fcos^2\theta _c} {(2\pi )^3}TeBE^2_\nu \left[(C^2_V + C^2_A)(1 + 2cos^2\theta ) - 6C_VC_Acos\theta \right]
\end{equation}
for the untrapped case.
\end{subsection}


\vskip .5 cm
\begin{subsection}{Non-degenerate matter}
\noindent Integrals can again be performed in the non-degenerate limit and 
we get

\begin{eqnarray}
\frac {\sigma _{A}(E_{\nu},B)} {V} 
&\simeq & 
	\frac {G^2_Fcos^2\theta_c} {4\pi }eB \cos_{\theta}n_{N}
	\frac{1}{e^{-(E_{\nu}+Q-mu_{e})\beta}+1}\nonumber \\
&&{}	[((g_V + g_A)^2 + 4g^2_A) + ((g_V + g_A)^2 - 4g^2_A)]\nonumber \\
\end{eqnarray}

\begin{eqnarray}
\frac {\sigma _{\nu p}(E_\nu ,B)} {V} 
&\simeq  &
	\frac {2G^2_F\cos_{\theta_{c}}^{2}C^2_A} {(2\pi )^3}eB
	\int _{0}^{\infty }dp_{2z}f_2(E_2) (E_\nu +
	\frac {p^2_{2z}} {2m_p} )\nonumber \\
&&{}	[ (E_\nu + \frac {p^2_{2z}} {2m_p} ) (2-\frac {1} {3m_p} 
	(E_\nu + \frac {p^2_{2z}} {2m_p}) ) 
	- \frac {E^2_\nu cos^2\theta + p^2_{2z}} {m_p} ]\nonumber \\
&&{}	\simeq \frac {2G^2_F\cos_{\theta_{c}}^{2}C^2_A} {2\pi }E^2_\nu n_p\nonumber \\
\end{eqnarray}
where
\begin{equation}
n_p = \frac {eB} {2\pi ^2}\int f_2(E_2)dp_{2z}
\end{equation}
and we have ignored terms of $O\left (\frac {p^2} {2m_p}\right )$.
\par
For completeness, we give below the neutrino-electron cross-section 
for the case of non-degenerate electrons too though, here we do not 
consider the regime of non-degenerate electrons.

\begin{eqnarray}
\frac {\sigma _{\nu e}(E_\nu ,B)} {V} 
& \simeq &
	\frac {4G^2_Fcos^2\theta_c} {(2\pi )^3}eB
	[(C^2_V + C^2_A)(4 + cos^2\theta)\nonumber \\
&&{}	- 4C_VC_Acos \theta ] E_\nu \int_{0}^{\infty } 
	\frac {dE_eE_e} {1 + e^{\beta(E_e - \mu _e)} }\nonumber \\
\end{eqnarray}

\begin{equation}
\simeq \frac {G^2_F} {\pi}n_eTE_\nu\left [(C^2_V + C^2_A)(1 + cos^2\theta ) - 4C_VC_Acos\theta\right ]
\end{equation}
where
\begin{equation}
n_e = \frac {2eB} {(2\pi )^2}\int _{0}^{\infty }e^{-\beta p_{ez}}dp_{ez} = \frac {2eBT} {(2\pi)^2}
\end{equation}
for non-degenerate electrons.
\end{subsection}
\end{section}

\begin{section}{Results and Discussions}

\noindent We now present neutrino opacity results for interacting magnetised 
neutron star matter consisting of nucleons, electrons and trapped as well as freely streaming neutrinos for densities ranging over a wide range 
$(10^{-5}-10)n_{0}$, temperature (5-60)MeV and magnetic field varying from 
$(0-5\times 10^5)B_{e}^c$. The composition of matter at these conditions 
has already been shown in figures (1-3) in Section 2 and the effect 
of magnetic field including anomalous magnetic moment of nucleons discussed.
We first consider matter in the core of neutron stars where matter 
is degenerate. At these densities relativistic effects can be 
important and should not be neglected, however, for simplicity we have
evaluated the matrix elements in the limit of non-relativistic nucleons.
Relativistic kinematic effects are expected to be typically of the order 
$\frac{p_{F}}{M^{\star}}$ in the degenerate limit and one can modify the 
limiting expressions (27-29,43,46) for absorption and scattering 
cross-sections by replacing the effective mass of the nucleons 
$(M_{n}^{\star},M_{p}^{\star})$ by the corresponding effective 
chemical potentials $(\mu_{n}^{\star},\mu_{p}^{\star})$ respectively 
(see ref. 1 for a discussion). In our numerical calculation of the neutrino
mean free path we have made this change. In the neutrino free case 
$(Y_{\nu_e}=0)$, the neutrino absorption cross-section due to reaction 
(1), the direct URCA process, is highly supressed due to simultaneous 
non-conservation of energy and momentum for degenerate matter. This happens
because for degenerate nuclear matter, the direct URCA process can take place
only near the Fermi energies of participating particles and simultaneous energy
momentun conservation sets the constraint 
$p_{F}(e)+p_{F}(p)\geq p_{F}(n)$, the proton fraction 
required for this costraint to be satisfied has to be greater than 
$\sim 11 \%$. The proton fraction calculation in nuclear matter is however
very sensitive to the assumption that go into the microscopic theory of
nuclear interactions which are not well known [15]. In the present work,
this constraint is satisfied at moderately high densities namely
$n_{B}>1.5 n_{0}$ (see also [11,16]), whereas in some recent potential
models with three-body interactions and boost corrections, this occurs at 
densities in excess of $\sim 4.5 n_{0}$ [17].Since the effect of 
magnetic field is to increase the proton fraction, at sufficiently high 
magnetic fields depending on the density, the constraint is satisfied and 
direct URCA process proceeds. In the case of extremely high magnetic 
fields capable of confining electrons and protons into the lowest 
Landau level, this constraint is not even required and the absorption 
process proceeds as can be seen from equation (42). Furtermore, in this 
situation the cross-section develops anisotropy. Similarly at high 
temperatures (see figs. 1-2), the proton fraction rises and the 
contribution of the direct process exceeds the neutrino-nucleon 
scattering process.

\noindent In the neutrino trapped regime, all the particles are degenerate
and the charged current absorption process dominates over the scattering 
processes; the neutrino-nucleon scattering being the most important of the 
scattering processes. In fig. 4 we show the neutrino absorption mean free 
path as a function of density for different temperatures and magnetic
fields for degenerate matter. The upper and lower figures correspond to
neutrino free $(Y_{\nu_e}=0)$ and neutrino trapped $(Y_{L}=0.4)$ matter.
In the neutrino free regime, the neutrinos are thermal and we take 
the neutrino energy to be equal to $3T$ for calculating the mean free paths
whereas, in the trapped regime, the neutrinos being degenerate, 
$E_{\nu}=\mu_{\nu}$ is the appropriate neutrino energy. 
In this high dcensity regime, the effect of magnetic field on 
absorption mean free path is not significant unless one goes to extremely 
high magnetic fields where the matter becomes completely polarised 
and the lowest Landau energy level is occupied.    
This happens for $B \geq 5\times 10^5 B_{e}^{c}$
in this density regime. The rapid fluctuations in absorption mean free path
around the zero field value reflects the appearance of a new Landau level
contributing to the reaction rate.  
For the case of polarising magnetic field, we have evaluated the absorption 
mean free path for neutrinos propogating along the magnetic field.
The asymmetry in the absorption cross-section for neutrinos propogating
along and opposite to the direction of magnetic field is roughly of order of 
$10 \%$
In fig. 5 we exhibit the neutrino scattering mean free paths 
$\lambda_{n}, \lambda{p}, \lambda_{e}$ on neutrons, protons and electrons 
respectively for $B=0$ and $5\times 10^4$ and at T=5, 30, 60 MeV.
The left panel correspond to the neutrino free and the right panel to
the neutrino trapped degenerate matter.
The effect of magnetic field on these scattering mean free paths is not 
significant at these densities.
At low densities $n_{B}=(10^{-6}-10^{-2})n_{0}$, the
nucleons are non-degenerate, the electrons continue to remain 
degenerate except at high temperatures and the matter becomes neutrino free.
Magnetic field in this regime has important effect. 
In fig. 6 we have plotted the absorption mean free path as a function of 
density at T=5 MeV  for magnetic field varying between 
$0$ to $5\times 10^4$. We find that the absorption cross-section increases
i.e mean free path decrease with the increase in magnetic field.
In fig. 7 we have plotted the neutrino scattering mean free paths 
$\lambda_{n}, \lambda{p}, \lambda_{e}$ as a function of density at 
T=5, 30, 60 MeV for B=0 and $5\times 10^4 B_{e}^c$. We find that
whereas, $\lambda_{n}$ and  $\lambda_{e}$ increase with the magnetic field, 
$\lambda_{p}$ decreases.
 
\noindent In summary we have studied the neutrino interactions rates
through charged as well as neutral current weak interactions in hot, dense
magnetised, interacting nuclear matter over a wide range of densities and
magnetic fields. The effect of including anomalous magnetic momemt of 
nucleons on the structure of nuclear matter shows up only when the 
magnetic field exceeds $\sim 10^5 B_{e}^c$ and results in further 
enhancement of proton fraction at low densities and lowering
of effective nucleon mass at high densities. The effect of magnetic 
field on neutrino mean free paths is most pronounced at low densities and
results in substantial decrease in the neutrino absorption mean free
path in addition to developing anisotropy.
\par
\vskip 1 cm
\noindent Acknowledgements: We are thankful to Prof. J V Narlikar for 
providing facilities at IUCAA, Pune where a part of this work was done.
One of us A.G. is thakful to University Grants Commission, India for partial
financial support. 

\end{section}

\pagebreak
\appendix{\bf Appendix A}
\indent

For determining the composition of dense, hot, magnetised matter, 
we employ a relativistic mean field theoretical approach in 
which the baryons (protons and neutrons) interact via the exchange of 
$\sigma-\omega-\rho$ mesons 
in a constant uniform magnetic field. Following reference {\cite{Agoyal}}
the thermodynamic potential of the system can be written as

$$ \hskip 2 cm\Omega=-\frac{1}{2}m_{\omega}^{2}\omega_{0}^{2}
	-\frac{1}{2}m_{\rho}^{2}\rho_{0}^{2}
	+\frac{1}{2}m_{\sigma}^{2}\sigma^{2}+U(\sigma)+\frac{B^2}{8\pi}
	+\sum_{i}\Omega_{i} \hskip 2 cm (A1)$$     

where $i=n,p,e, \nu$.

and 
$$\hskip 2 cm U(\sigma)=\frac{1}{3}bm_{N}(g_{\sigma_{n}}\sigma)^{3}+\frac{1}{4}c(g_{\sigma_{N}}\sigma)^{4} \hskip 4.5 cm (A2)$$

and 
$$\hskip 2 cm \Omega_{e}=-\frac{1}{\beta(2\pi)^{2}}\sum_{s} \sum_{\nu} \int 
	dp_{z}\ln[1+e^{-\beta ({\cal E}_{\nu,s}^{e}
	-\mu_{e})}] \hskip 3 cm (A3)$$

$$\hskip 2 cm \Omega_{p}=-\frac{1}{\beta(2\pi)^{2}}\sum _{s}\sum_{\nu}\int 
	dp_{z}\ln[1+e^{-\beta({\cal E}_{\nu,s}^{p}
	-\mu_{p}^{\star})}] \hskip 3 cm (A4)$$

$$\hskip 2 cm \Omega_{n}=-\frac{1}{\beta(2\pi)^{2}}\sum_{s}\int 
	d^3p\ln[1+e^{-\beta({\cal E}_{s}^{n}
	-\mu_{n}^{\star})}]   \hskip 3.5 cm (A5)$$

$$\hskip 2 cm \Omega_{\nu}=-\frac{1}{\beta(2\pi)^{2}}\int 
	d^3p\ln[1+e^{-\beta({\cal E}_{\nu} 
	-\mu_{\nu})}] \hskip 4 cm (A6) $$

The energy eigenvalues are given in equations (8-10). The chemical potentials 
$\mu_{N}^{\star}$ and the effective masses $m_{n}^{\star}$ of nucleons 
are given by

$$\hskip 3 cm \mu_{N}^{\star}=\mu_{N}-U_{0}^{N}\hskip 6 cm (A7) $$
and
$$\hskip 3 cm m_{N}^{\star}=m_{N}-g_{\sigma_{N}}\sigma_{0} \hskip 6 cm (A8) $$

$$\hskip 2 cm U_{0}^{N}=g_{\omega N}\omega_{0}+g_{\rho N}\tau_{3 N}\rho_{0}\hskip 5 cm (A9)$$

In the mean field approximation, the thermodynamic quantities are 
expressed in terms of thermodynamic averages of meson fields which 
are assumed to be constant and are related to the baryonic and scalar number densities through the fixed equations viz.

$$\hskip 2 cm m_{\sigma}^{2}<\sigma>+\frac{\partial U(\sigma)}{\partial\sigma}=g_{\sigma N}
	(n_{n}^{s}+n_{p}^{s})\hskip 4 cm (A10) $$

$$\hskip 2 cm m_{\omega}^{2}<\omega_{0}>=g_{\omega N}(n_{n}+n_{p})\hskip 5 cm (A11)$$
  
$$\hskip 2 cmm_{\rho}^{2}<\rho_{0}>=g_{\rho N}(n_{p}-n_{n})\hskip 5 cm (A12)$$

The number densities being given by

$$\hskip 2 cm n_{p}=\frac{eB}{2\pi^{2}}\sum_{s}\sum_{\nu}\int_{0}^{\infty} 
	\frac{dp_{z}}{1+e^{\beta({\cal E}_{\nu,s}^{p}-\mu_{p}^{\star})}}\hskip 3.5  cm (A13) $$

$$\hskip 2 cm n_{n}=\frac{1}{(2\pi)^{3}}\sum_{s}\int 
	\frac{d^3p}{1+e^{\beta({\cal E}_{s}^{n}-\mu_{n}^{\star})}}\hskip 4 cm (A14) $$

$$\hskip 2 cm n_{p}^{s}=\frac{eB}{2\pi^{2}}m_{p}^{\star}\sum_{s}\sum_{\nu}\int 
	\frac{dp_{z}}{{\cal E}_{\nu,s}^{p}(1+e^{\beta({\cal E}_{\nu,s}^{p}-\mu_{p}^{\star})})}\hskip 2  cm (A15)$$

$$\hskip 2 cm n_{n}^{s}=\frac{m_{n}^{\star}}{(2\pi)^{3}}\int 
	\frac{d^3p}{{\cal E}_{s}^{n}
	(1+e^{\beta({\cal E}_{s}^{n}-\mu_{n}^{\star})})}\hskip 4  cm (A16) $$

The net electron and neutrino number densities are given by

$$\bar n_{e}=\frac{eB}{2\pi^{2}}\sum(2-\delta_{\nu,0})\int 
	dp_{z}[\frac{1}{(1+e^(\beta({\cal E}_{\nu}^{e}-\mu_{e})))}
	-\mu_{e}\leftrightarrow(-\mu_{e})]\hskip 1.5 cm (A17) $$
and

$$\bar n_{\nu_{e}}=\frac{1}{(2\pi)^{3}}\int 
	d^3p[\frac{1}{(1+e^(\beta({\cal E}^{\nu}
	-\mu_{\nu})))}-\mu_{\nu}\leftrightarrow(-\mu_{\nu})]\hskip 2 cm (A18)$$

All the thermodynamic quantities are now obtained by solving the field 
equations ((A1)-(A18)) along with the condition of charge neutrality

$$\hskip 2 cm n_{p}=\bar n_{e}\hskip 6 cm (A19)$$
and trapped lepton fraction
$$\hskip 2 cm Y_{L}=Y_{e}+Y_{\nu_{e}}\hskip 5 cm (A20)$$
where 
$Y_{e}=\frac{\bar n_{e}}{n_{B}}$, $Y_{\nu_e}=\frac{\bar n_{ \nu_{e}}}{n_{B}}$
and the condition of $\beta$ equilibrium
$$\hskip 2 cm \mu_{n}=\mu_{p}+\mu_{e}-\mu_{\nu}\hskip 5 cm (A21)$$

self consistently for a given baryon density
$$\hskip 2 cm n_{B}=n_{p}+m_{n}\hskip 6 cm (A22)$$

The coupling constants $g_{\sigma N}$, $g_{\omega N}$, $g_{\rho N}$, b and c
are taken from Glendenning and Moszkowski {\cite{Glendenning}} by reproducing the nuclear
matter properties and the numerical values are given by

$$\frac{g_{\sigma N}}{m_{\sigma}}=3.434 fm^{-1}$$
$$ \frac{g_{\omega N}}{m_{\omega}}=2.694 fm^{-1} $$
$$ \frac{g_{\rho N}}{m_{\rho}}=2.1  fm^{-1}$$
$$ b=0.00295 $$
$$\hskip 4.8 cm c=-0.00107 \hskip 4  cm (A23)$$


\pagestyle{empty}
\begin{figure}[ht]
\centerline{ 
\epsfxsize=10cm\epsfysize=20cm\epsfbox{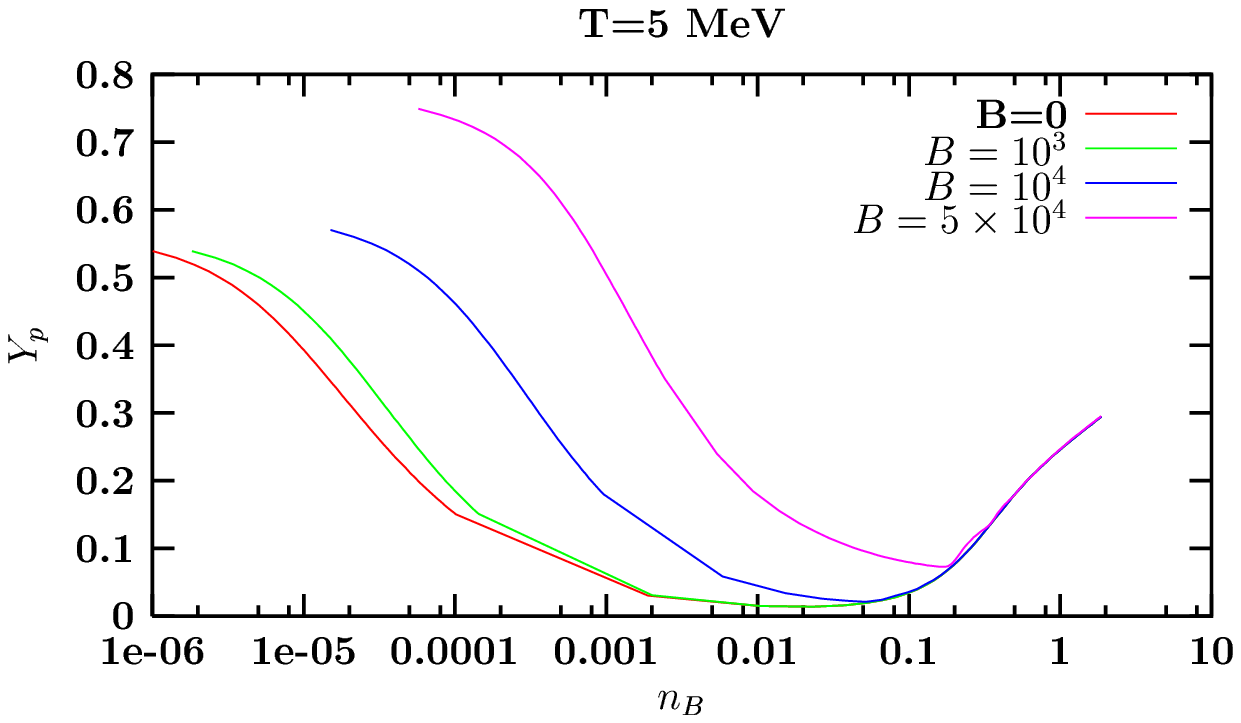}
	\hspace*{-2 cm}
\epsfxsize=10cm\epsfysize=20cm\epsfbox{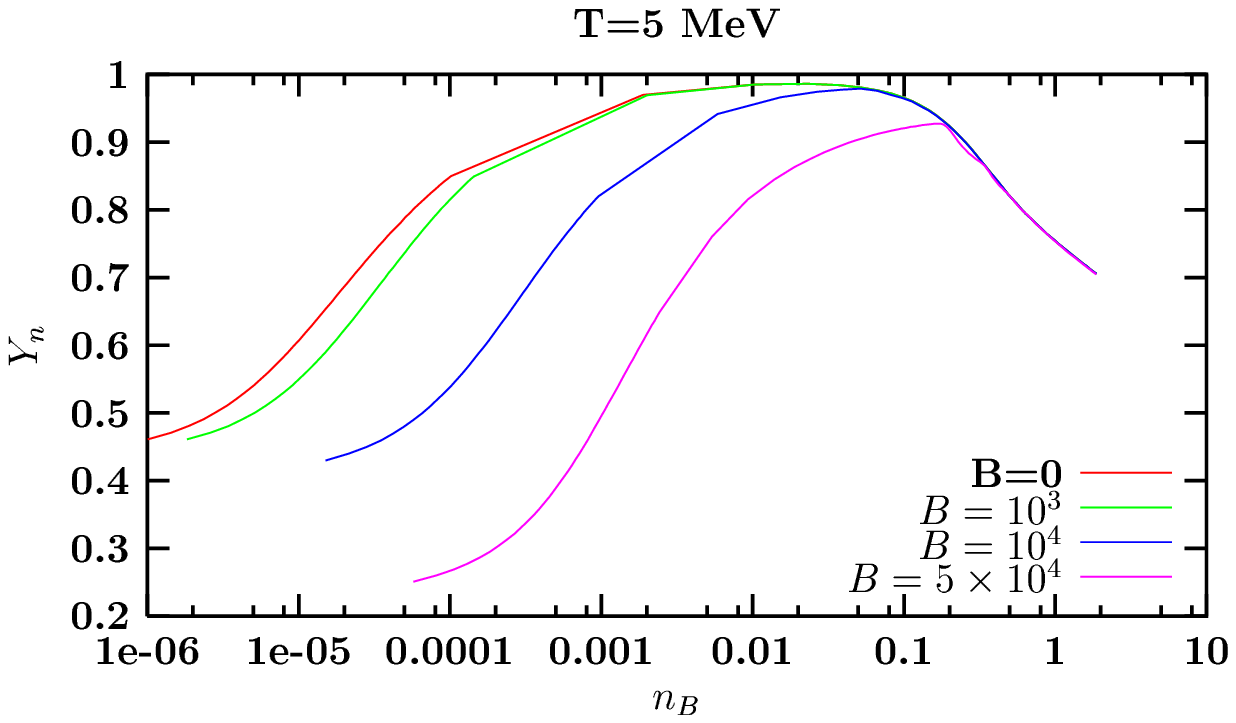}
	\hspace*{5 cm}
}
\vskip -14.5cm
\centerline{
\epsfxsize=10cm\epsfysize=20cm
                     \epsfbox{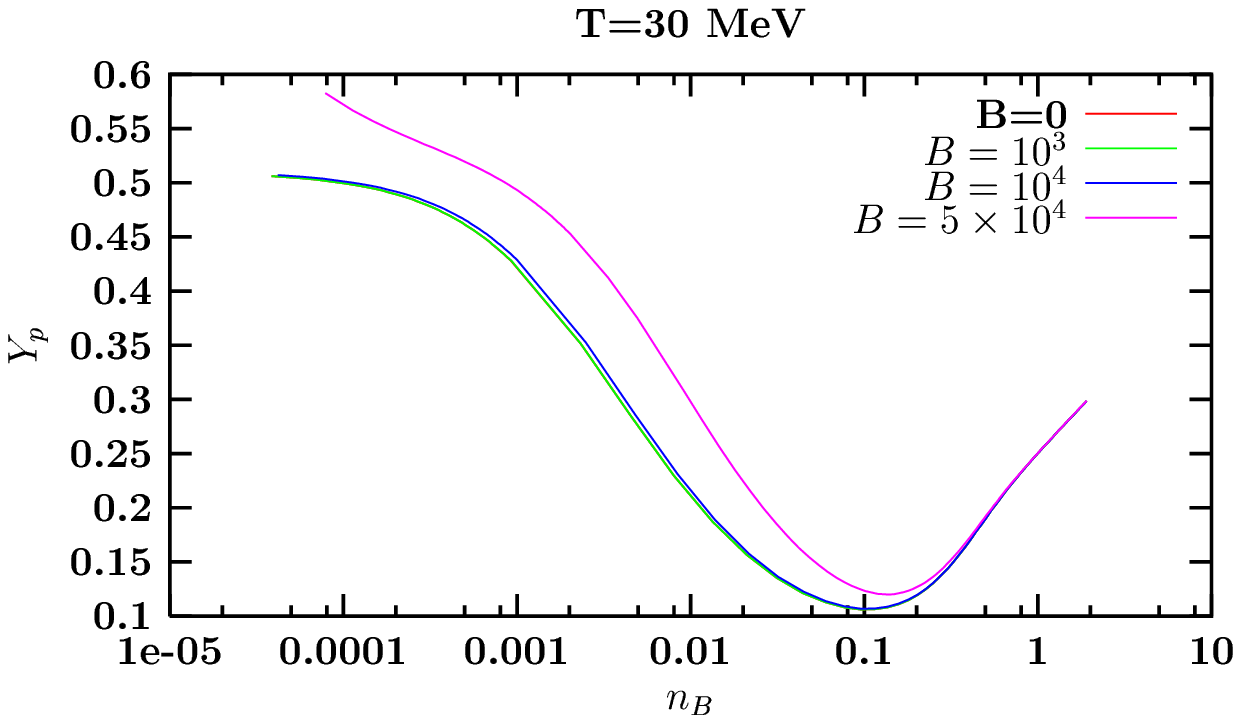}
	\hspace*{-2 cm}
\epsfxsize=10cm\epsfysize=20cm
                     \epsfbox{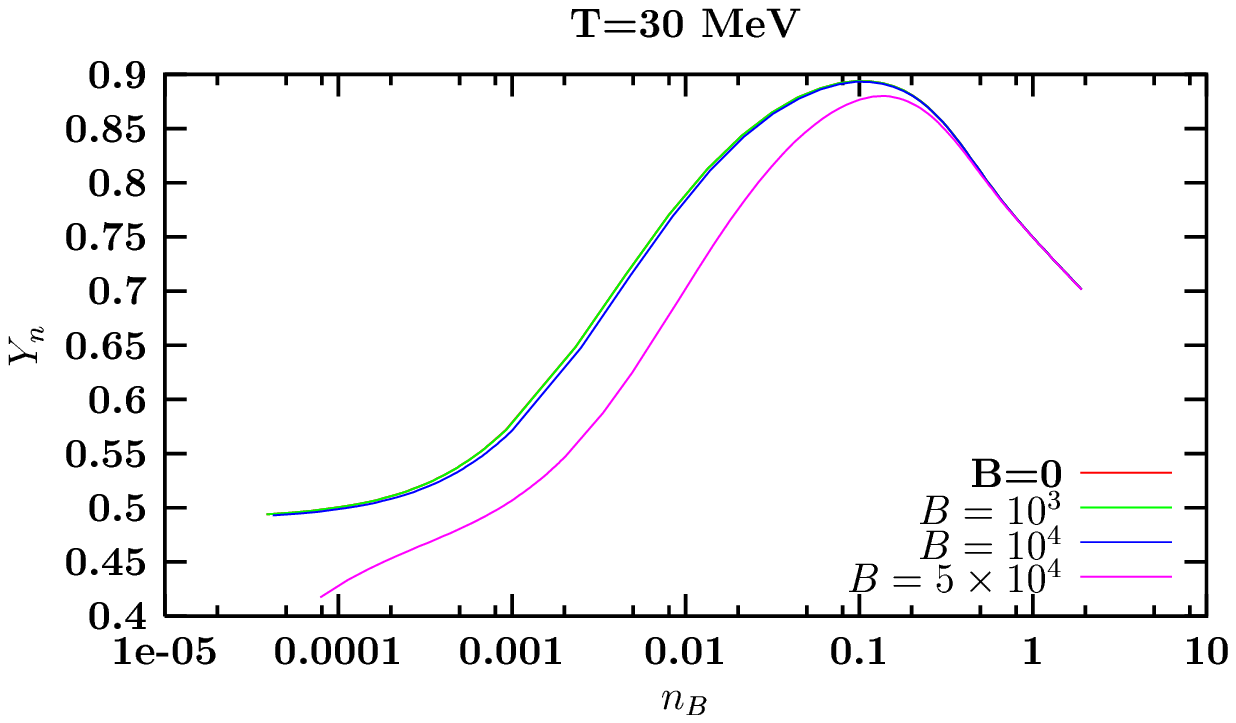}
	\hspace*{5 cm}
}
\vskip -14.5cm
\centerline{
\epsfxsize=10cm\epsfysize=20cm
		\epsfbox{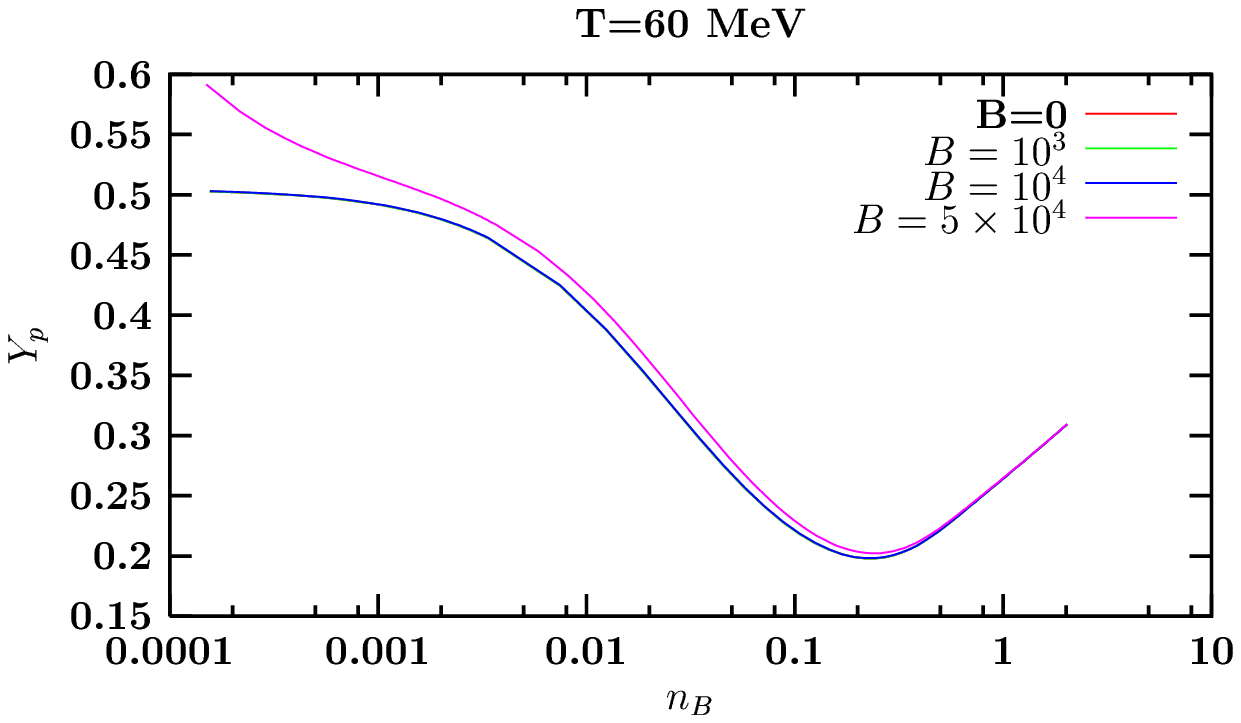}
	\hspace*{-2 cm}
\epsfxsize=10cm\epsfysize=20cm\epsfbox{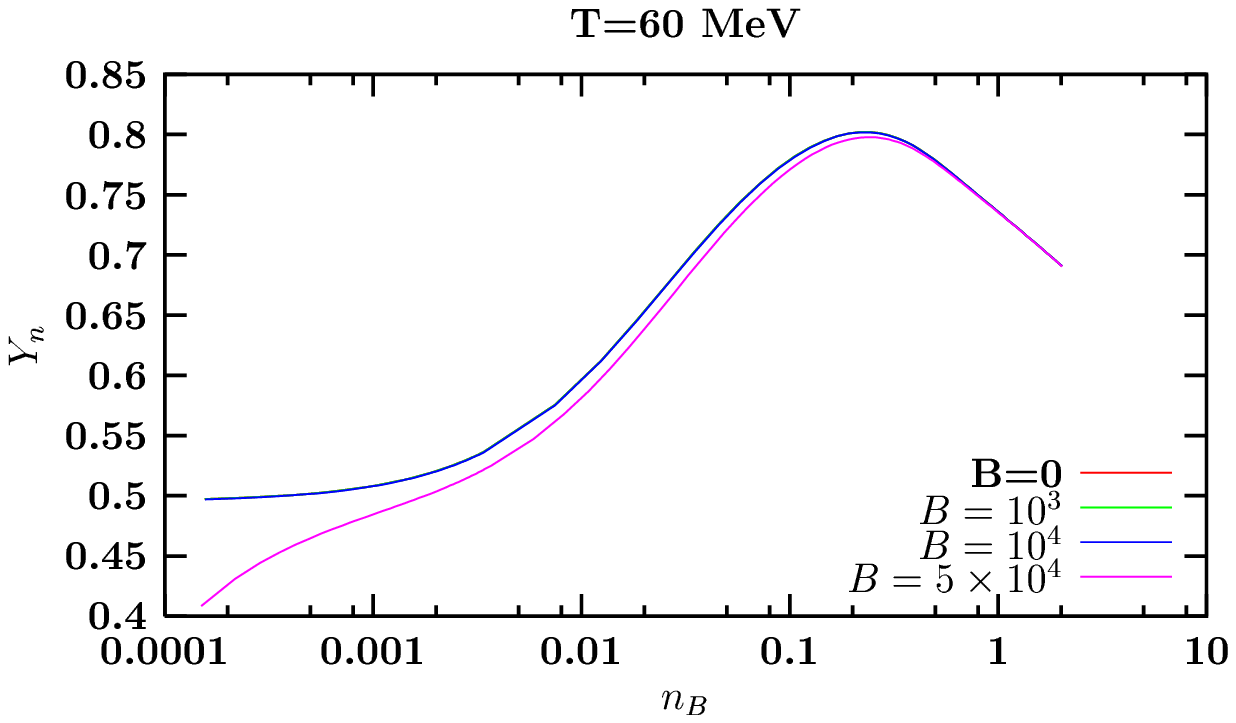}
	\hspace*{5 cm}
}
\hskip -8cm
\vskip -10cm

\caption{{\em Variation of $Y_{p}$ and $Y_{n}$ with $n_{B}$ for B=0, $10^3$, $10^{4}$ 
and $5\times 10^{4}$ for T=5, 30, 60 MeV ($Y_{\nu_ e}=0$).
}}
\end{figure}

\pagebreak
\begin{figure}[ht]
\centerline{ 
\epsfxsize=7cm\epsfysize=20cm\epsfbox{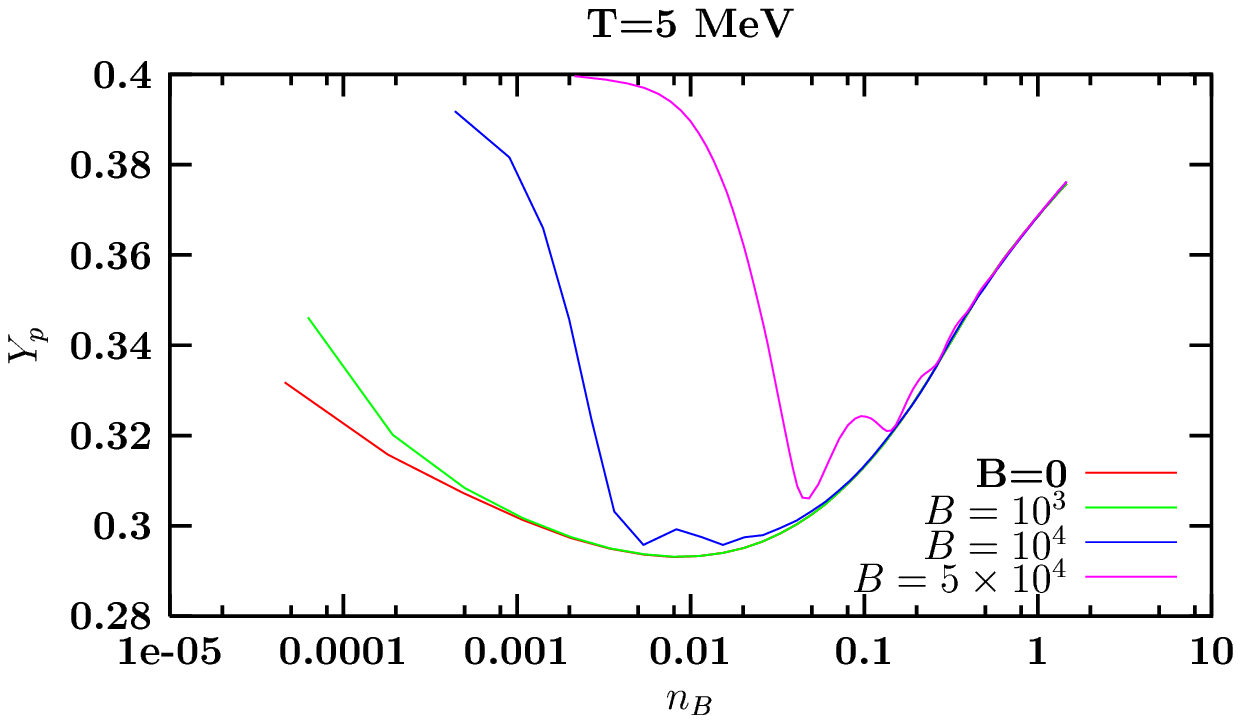}
	\hspace*{-2.5 cm}
\epsfxsize=7cm\epsfysize=20cm\epsfbox{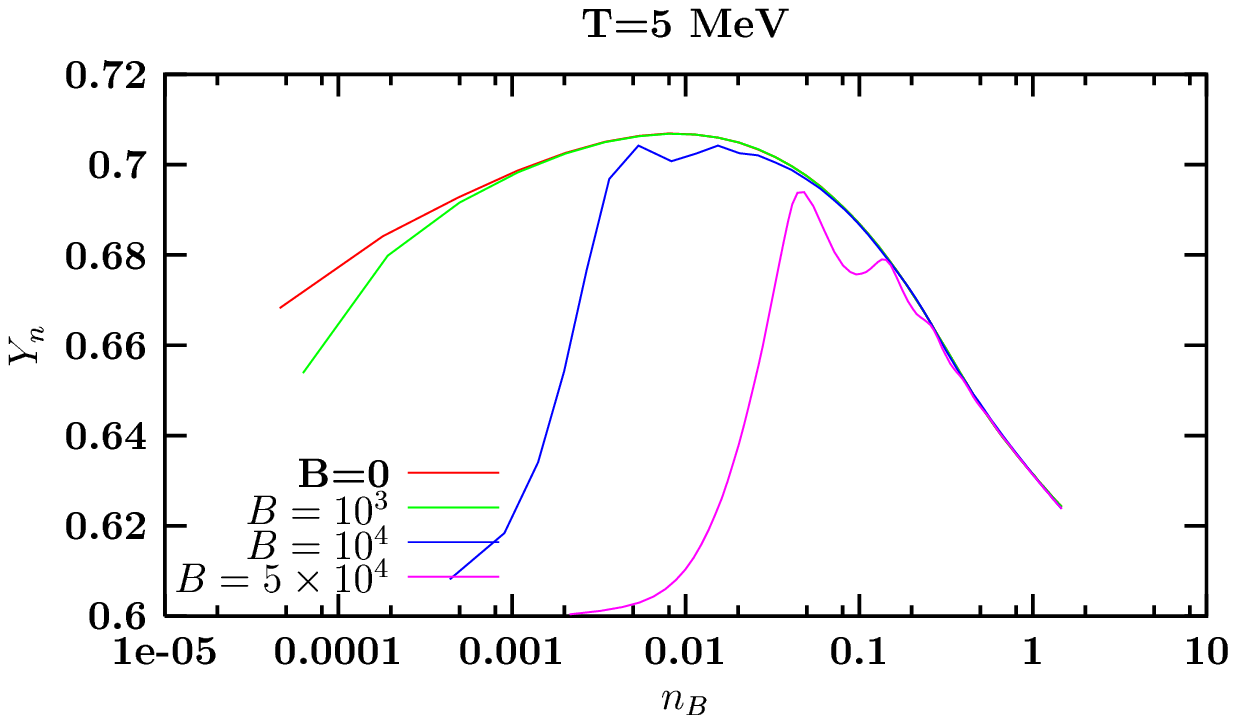}
	\hspace*{-2.5 cm}
\epsfxsize=7cm\epsfysize=20cm\epsfbox{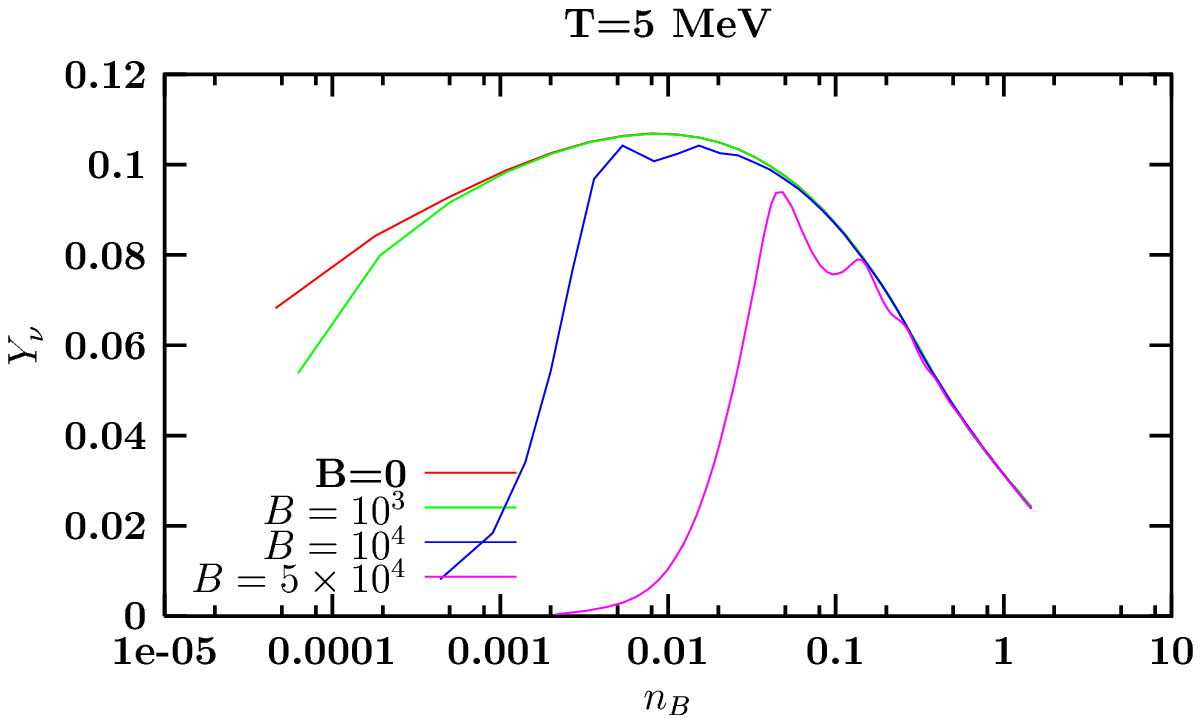}
	\hspace*{2 cm}
}
\vskip -14.5cm
\centerline{
\epsfxsize=7cm\epsfysize=20cm
                     \epsfbox{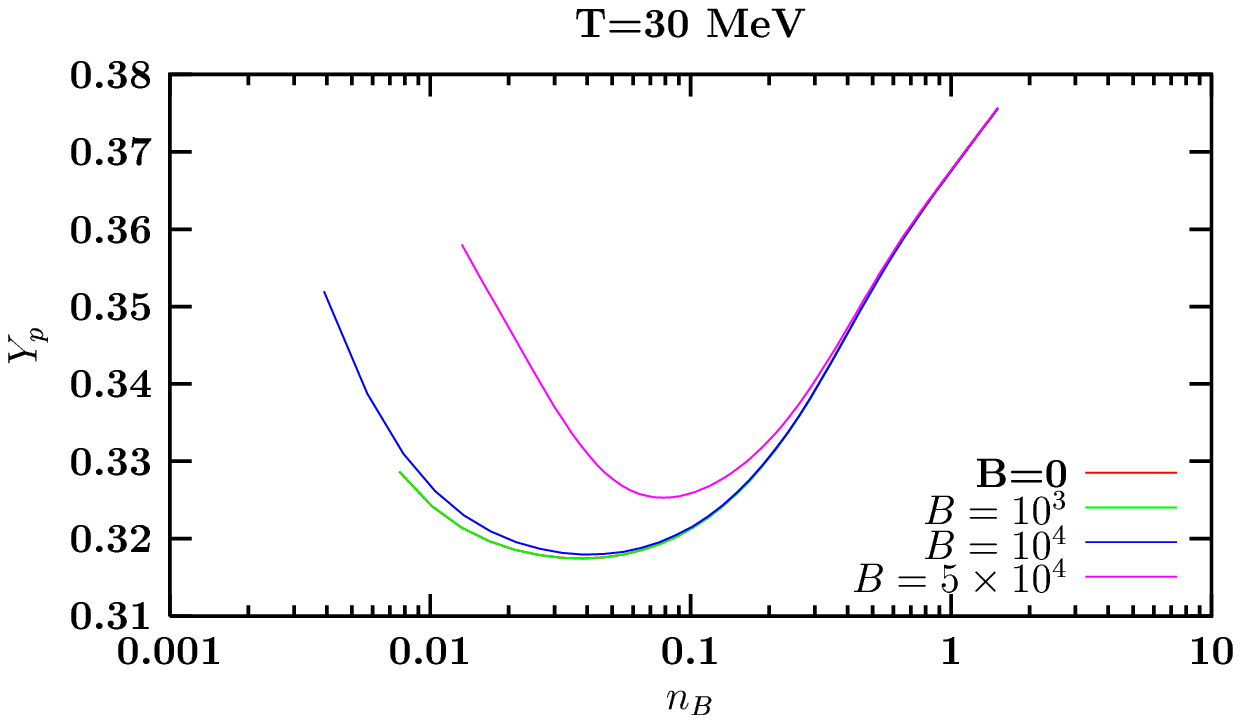}
	\hspace*{-2 cm}
\epsfxsize=7cm\epsfysize=20cm
                     \epsfbox{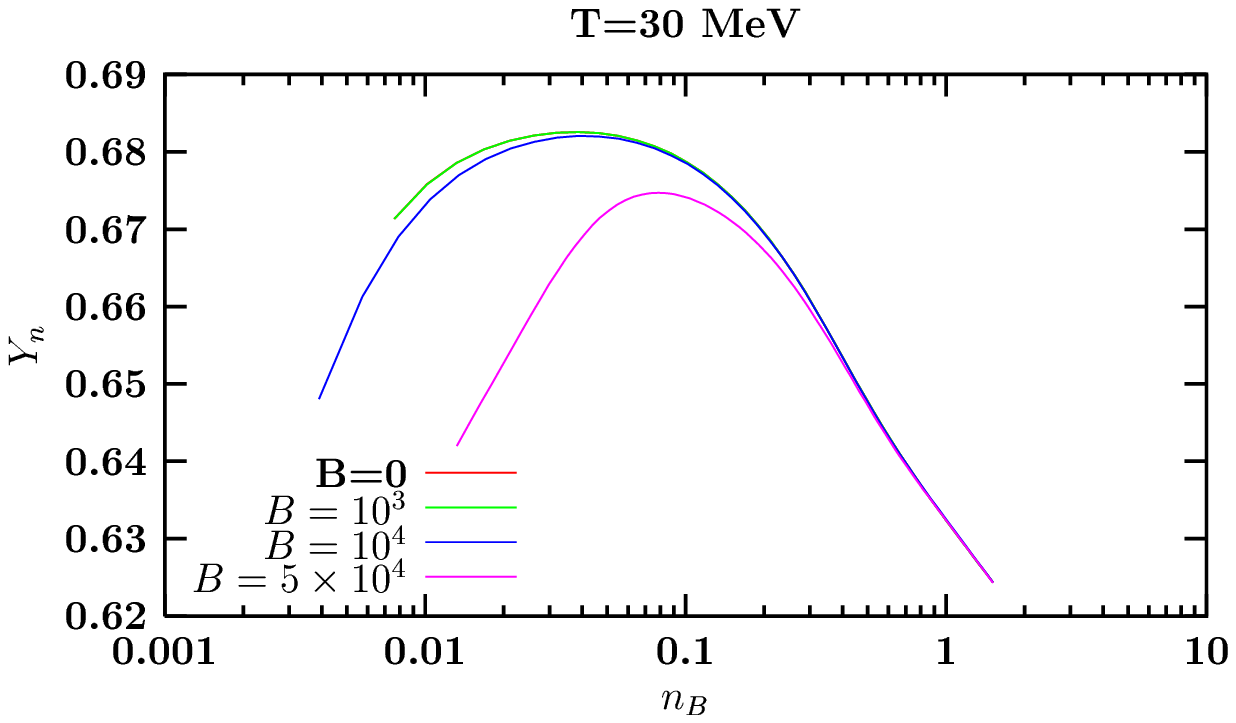}
	\hspace*{-2 cm}
\epsfxsize=7cm\epsfysize=20cm\epsfbox{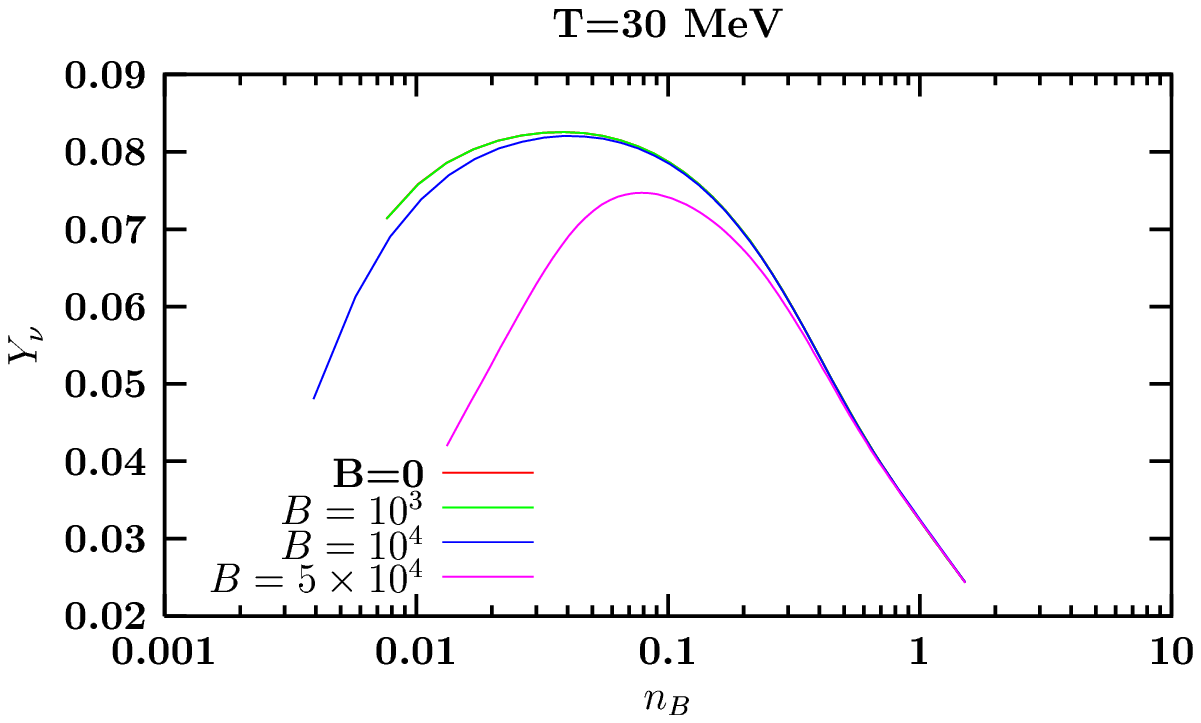}
	\hspace*{2 cm}
}
\vskip -14.5cm
\centerline{
\epsfxsize=7cm\epsfysize=20cm
		\epsfbox{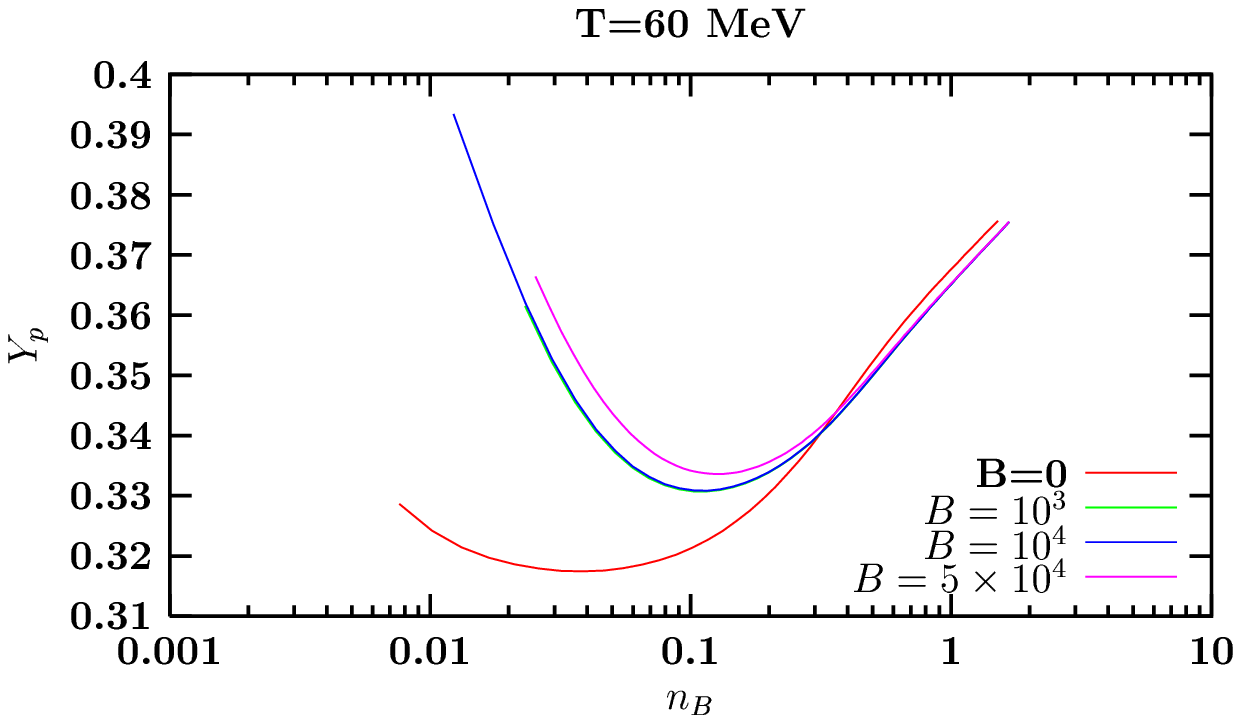}
	\hspace*{-2 cm}
\epsfxsize=7cm\epsfysize=20cm\epsfbox{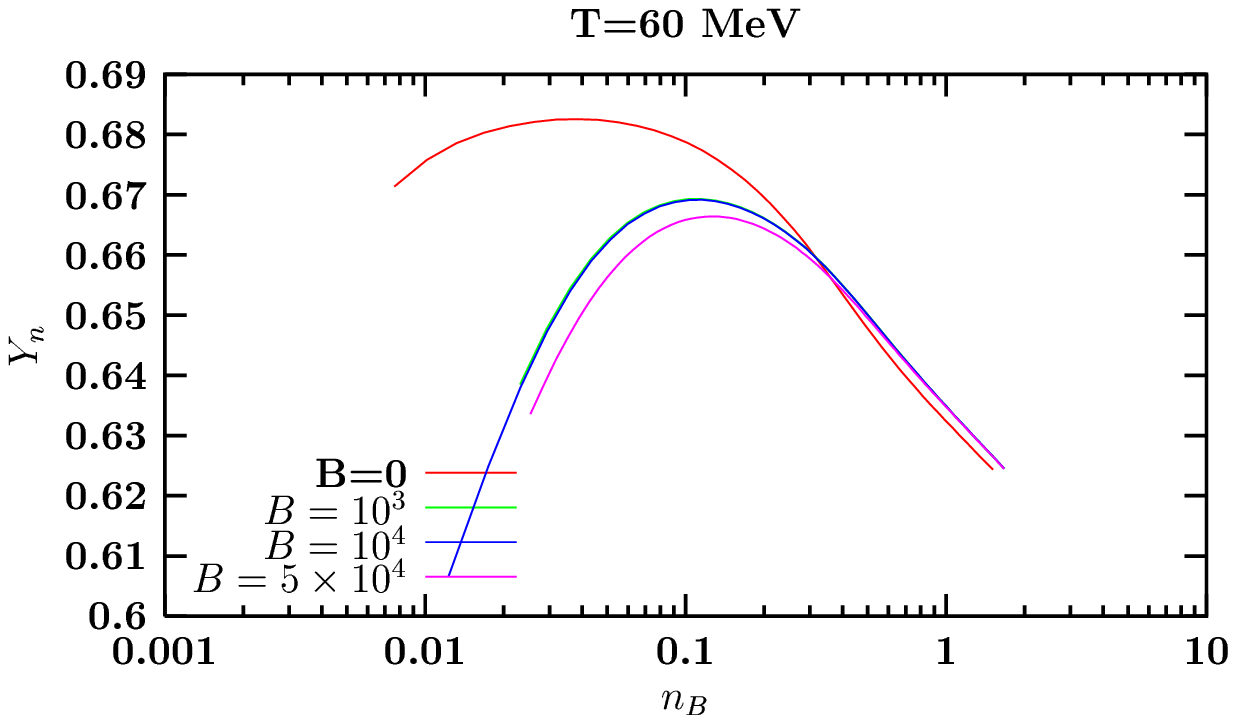}
	\hspace*{-2 cm}
\epsfxsize=7cm\epsfysize=20cm\epsfbox{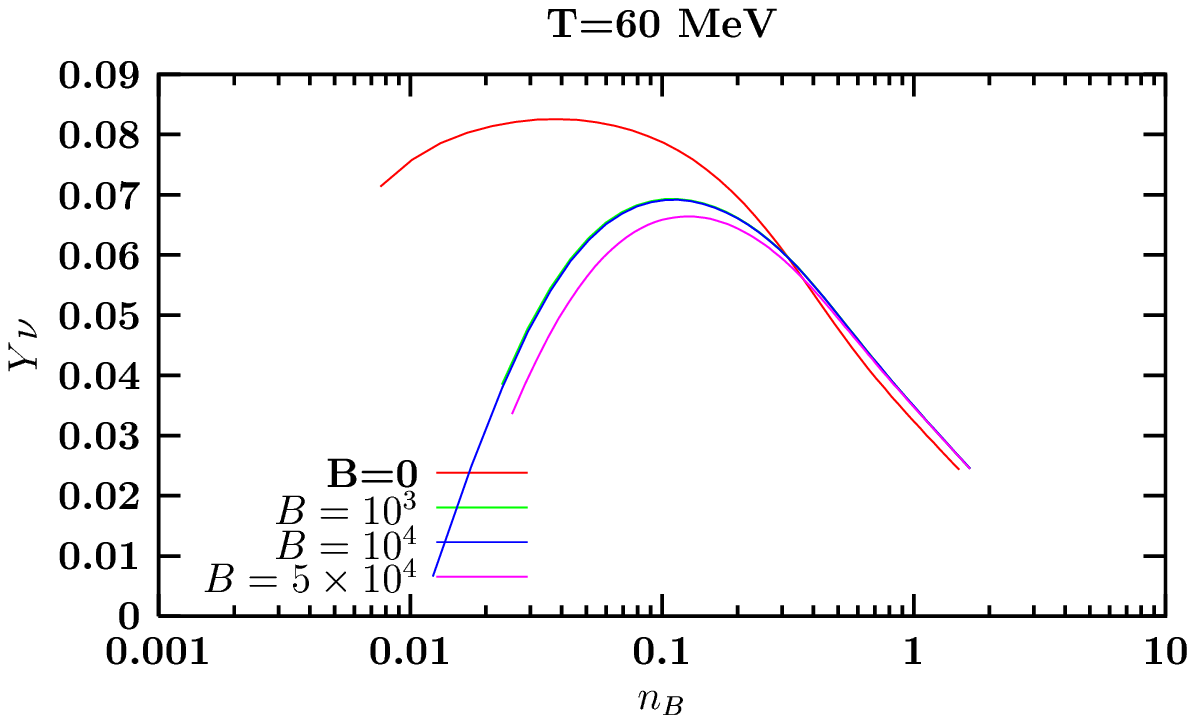}
	\hspace*{2 cm}
}
\hskip -8cm
\vskip -10cm

\caption{{\em Variation of $Y_{p}$, $Y_{n}$ and $Y_{\nu}$ with $n_{B}$ for B=0, $10^3$, $10^{4}$ 
and $5\times 10^{4}$ for T=5, 30, 60 MeV ($Y_{L}=0.4$).
}}

\end{figure}

\pagebreak
\begin{figure}[ht]
\begin{center}
{\Large {\bf T=5 MeV}}
\end{center}
\vskip -4cm
\centerline{ 
\epsfxsize=10cm\epsfysize=26cm\epsfbox{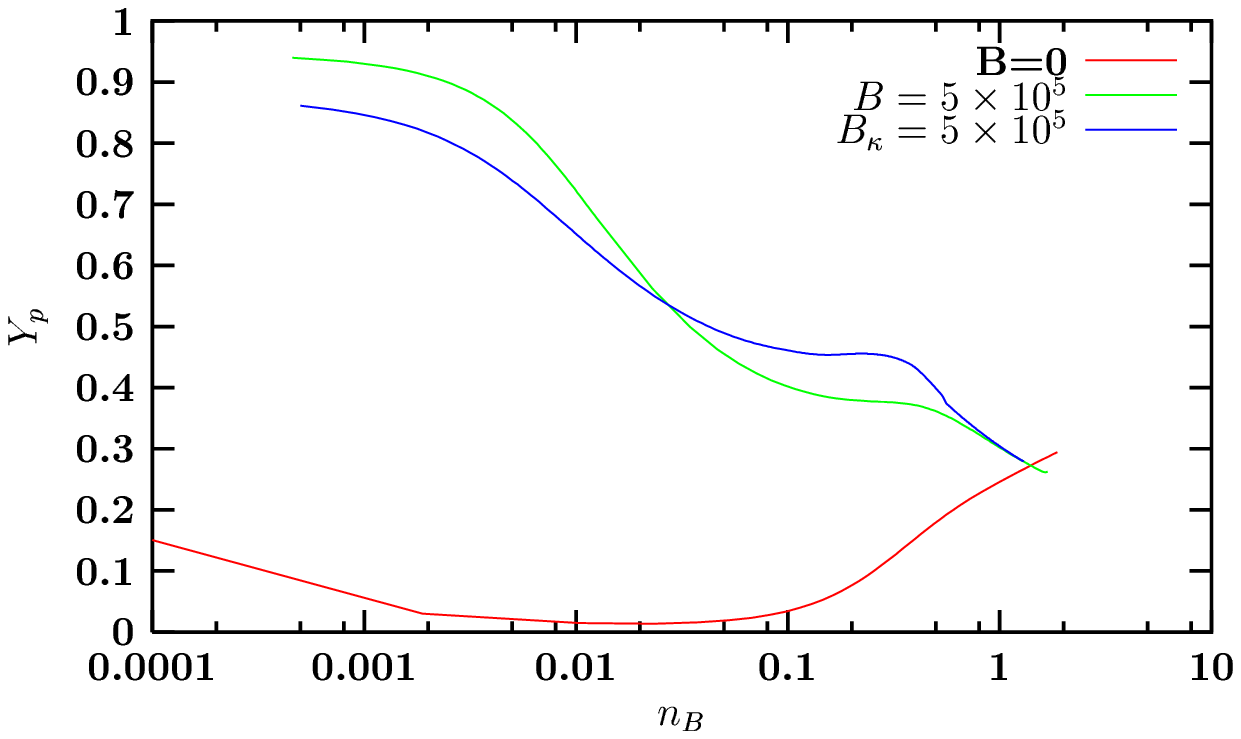}
	\hspace*{-4 cm}
\epsfxsize=10cm\epsfysize=26cm\epsfbox{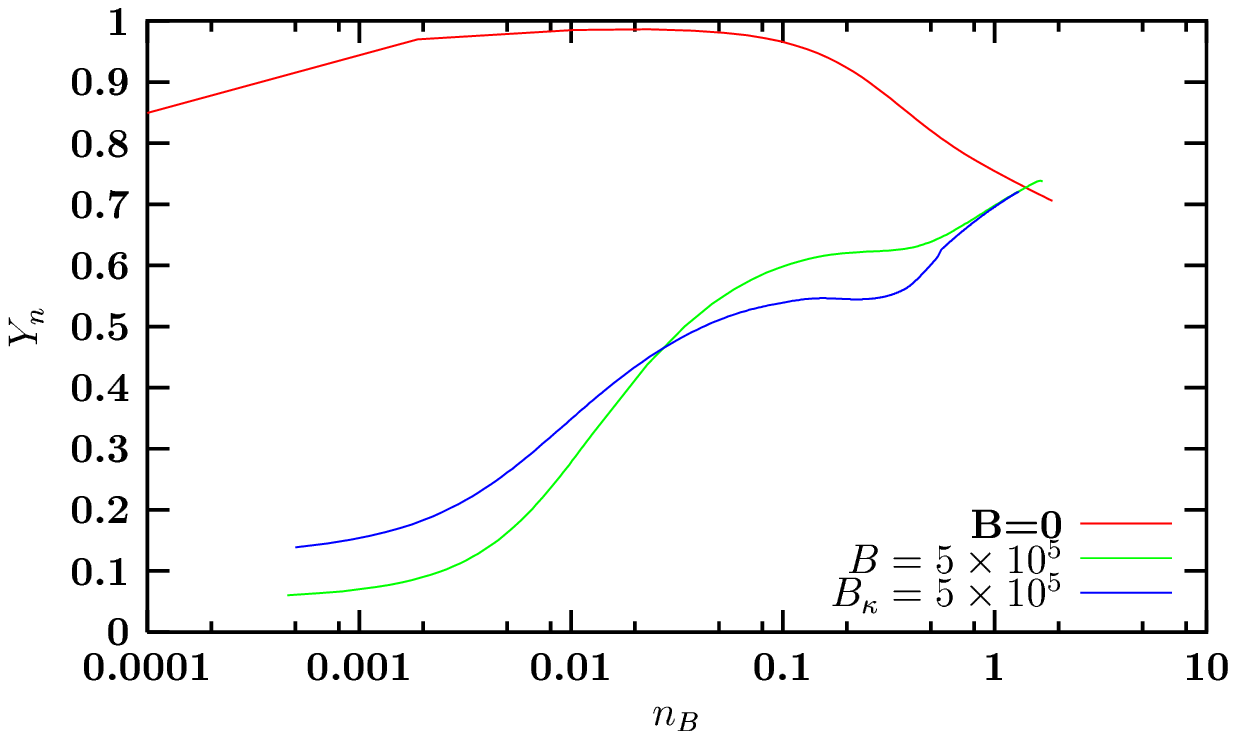}
}

\vskip -18.5cm
\centerline{ 
\epsfxsize=12cm\epsfysize=24cm\epsfbox{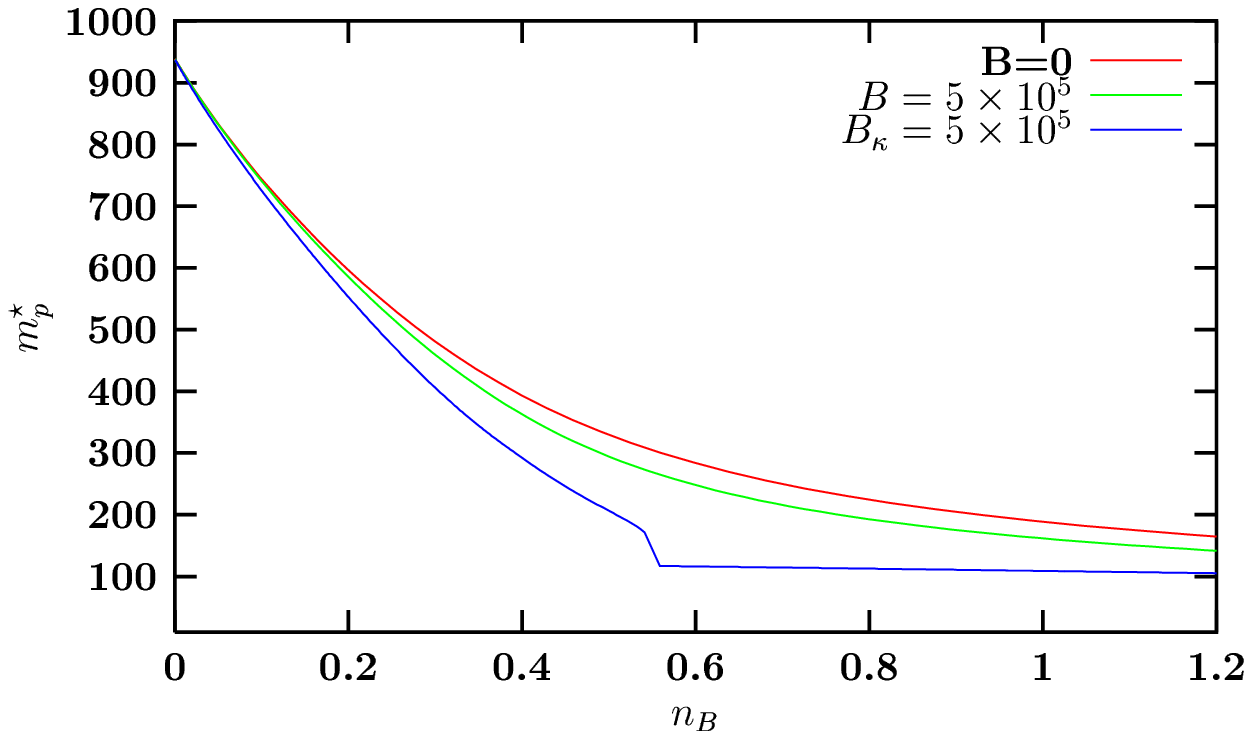}
}

\vskip -13.5cm
\caption{{\em Variation of $Y_{p}$, $Y_{n}$ and $m_{p}^\star$ with $n_{B}$ 
for $B=0, 5\times 10^5$ and $B_{\kappa}=5\times 10^5$ where in 
$B_{\kappa}$, the effect of anamalous magnetic has been included.   
}}

\end{figure}

\pagebreak
\begin{figure}[ht]
\centerline{ 
\epsfxsize=15cm\epsfysize=29cm\epsfbox{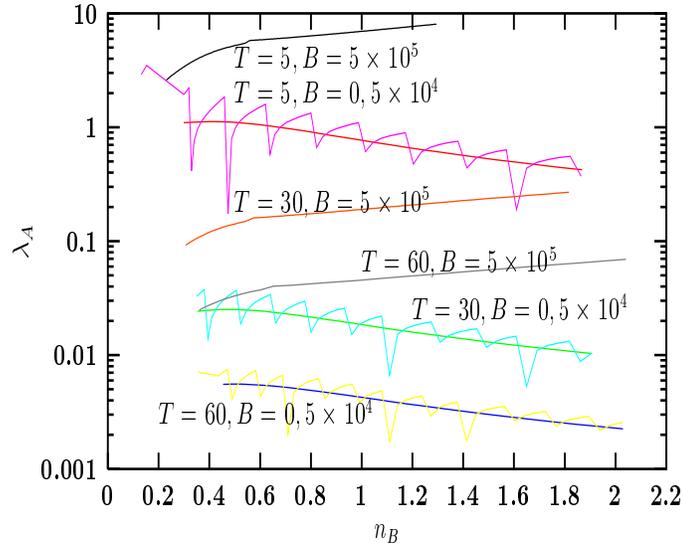}
}
\vskip -21cm

\centerline{ 
\epsfxsize=15cm\epsfysize=29cm\epsfbox{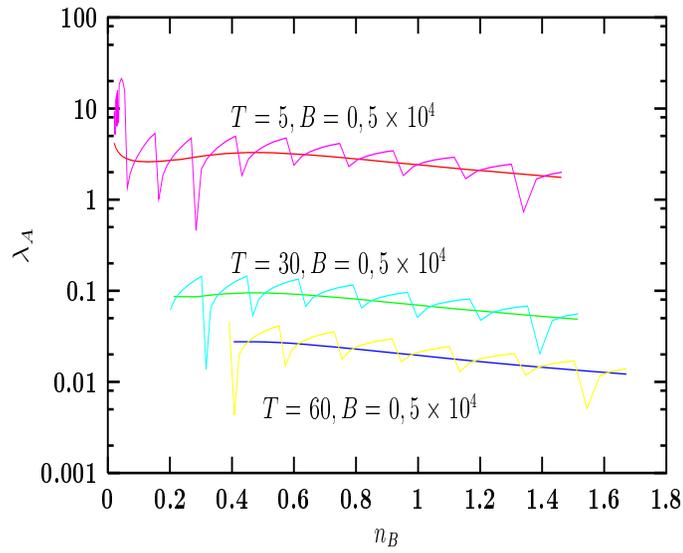}
}

\vskip -17cm
\caption{{\em The first figure shows the Variation of $\lambda_{A}$ 
with $n_{B}$ for untrapped degenrate matter for $B=0$, $5\times 10^4$ and 
$5\times 10^5$ at T=5, 30, 60 MeV. The second figure shows the variation of 
$\lambda_{A}$ with $n_{B}$ for trapped degenrate matter for $B=0$ and 
$5\times 10^4$ at T=5, 30, 60 MeV. The effect of anamalous 
magnetic moment has been included for all fields.   
}}

\end{figure}

\pagebreak
\begin{figure}[ht]
\centerline{ 
\epsfxsize=12cm\epsfysize=20cm\epsfbox{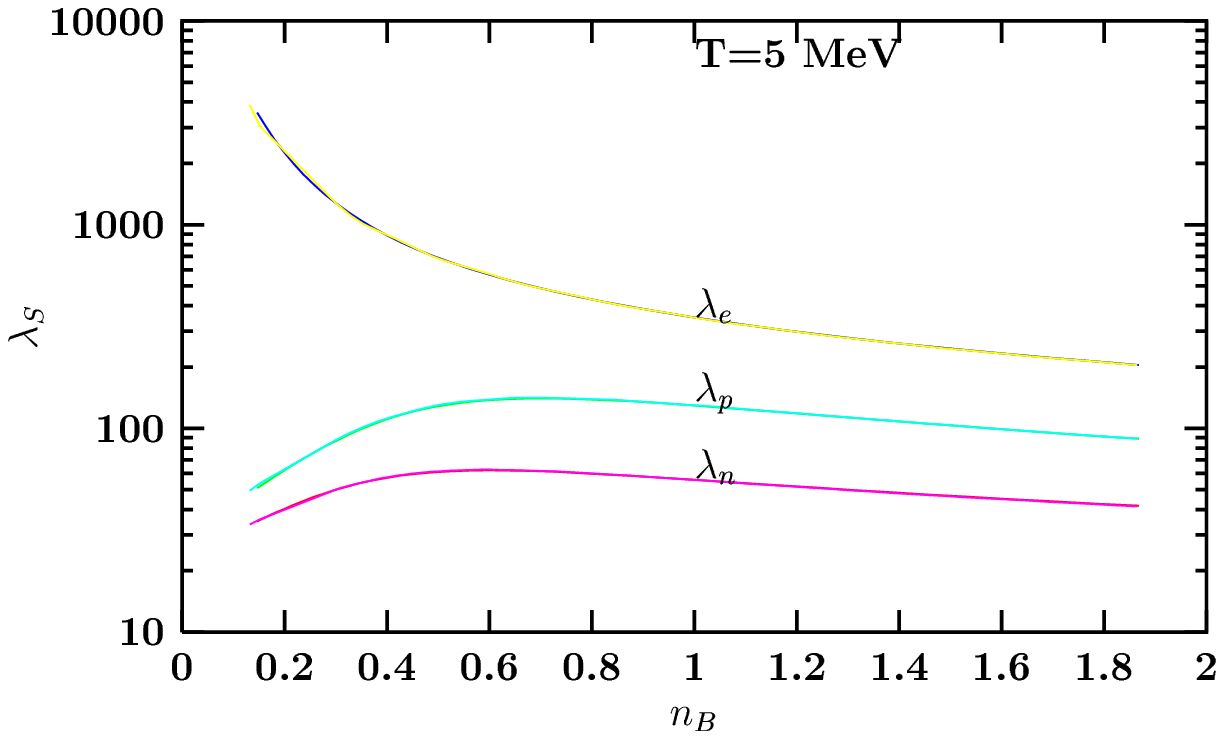}
	\hspace*{-4 cm}
\epsfxsize=12cm\epsfysize=20cm\epsfbox{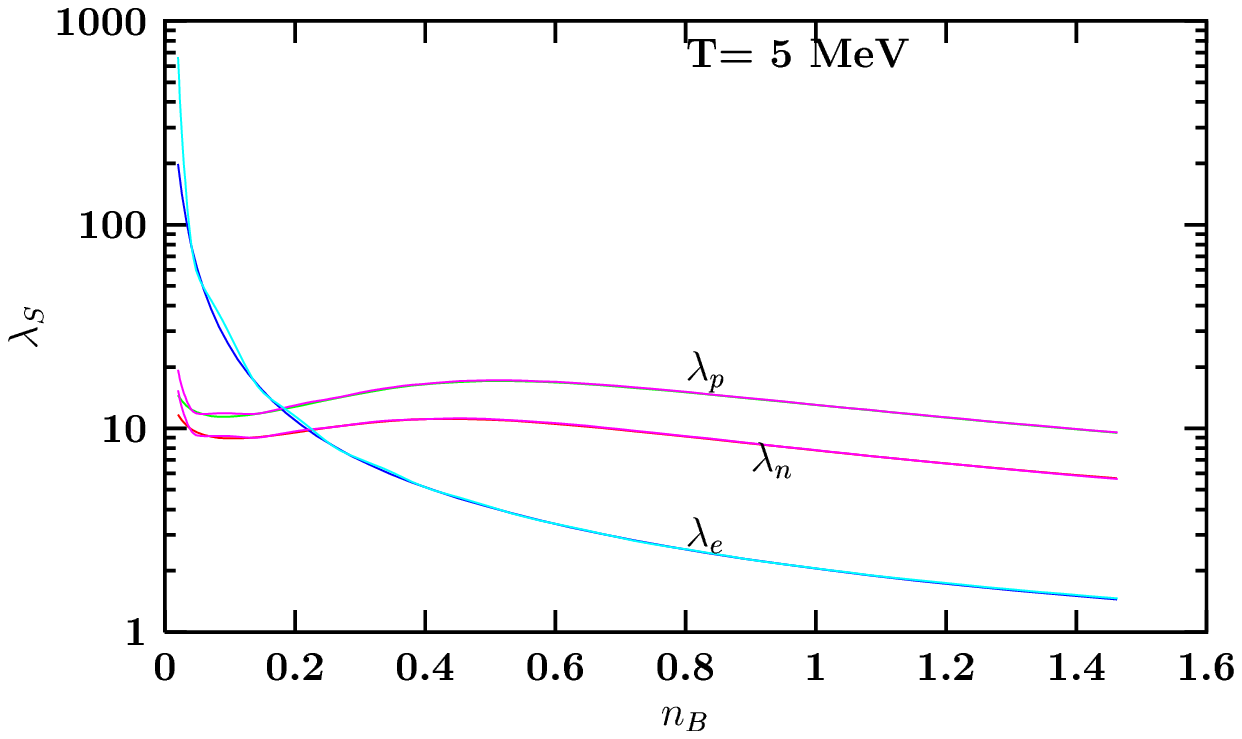}
	\hspace*{5 cm}
}
\vskip -14.5cm

\centerline{ 
\epsfxsize=12cm\epsfysize=20cm\epsfbox{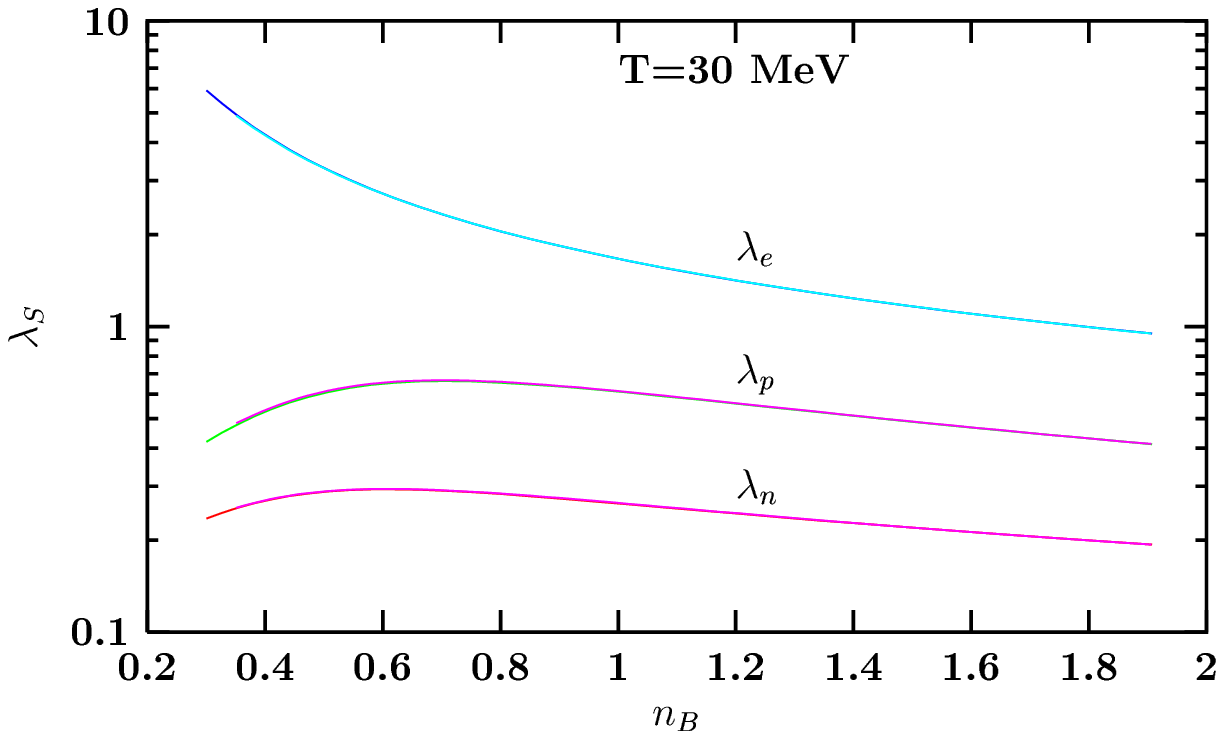}
	\hspace*{-4 cm}
\epsfxsize=12cm\epsfysize=20cm
                     \epsfbox{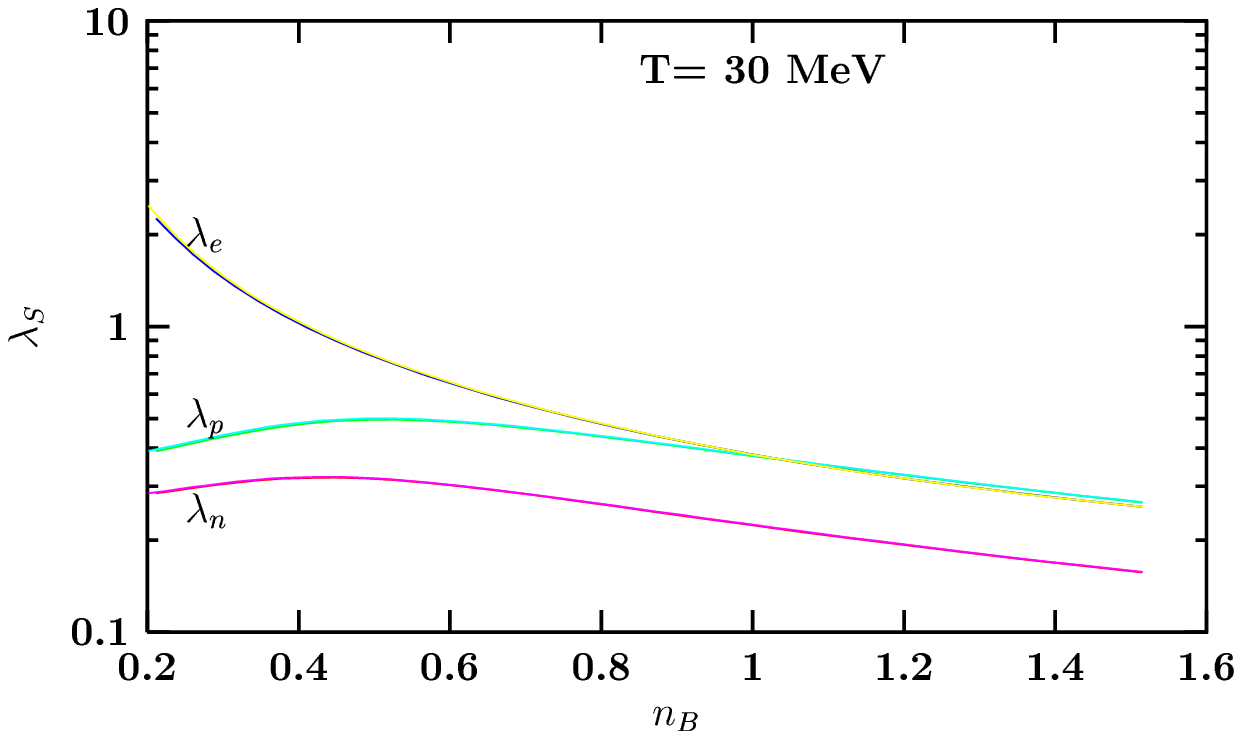}
	\hspace*{5 cm}
}

\vskip -14.5cm
\centerline{ 
\epsfxsize=12cm\epsfysize=20cm\epsfbox{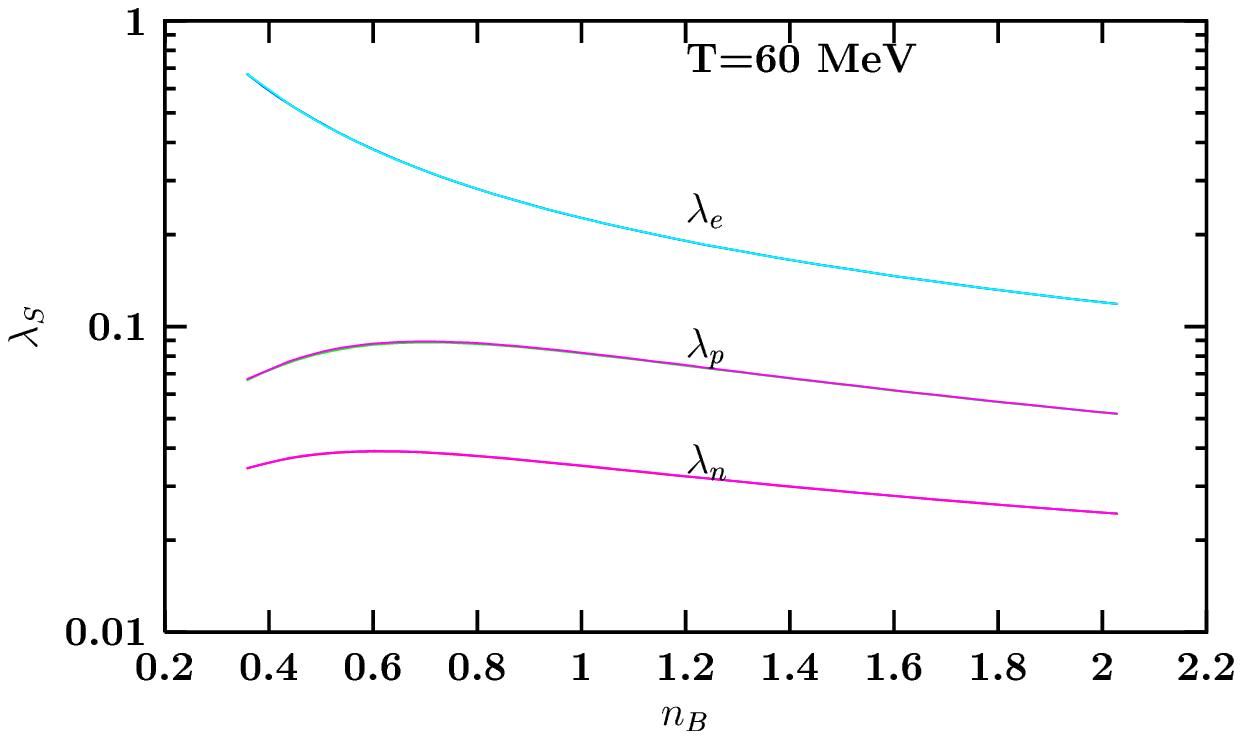}
	\hspace*{-4 cm}
\epsfxsize=12cm\epsfysize=20cm\epsfbox{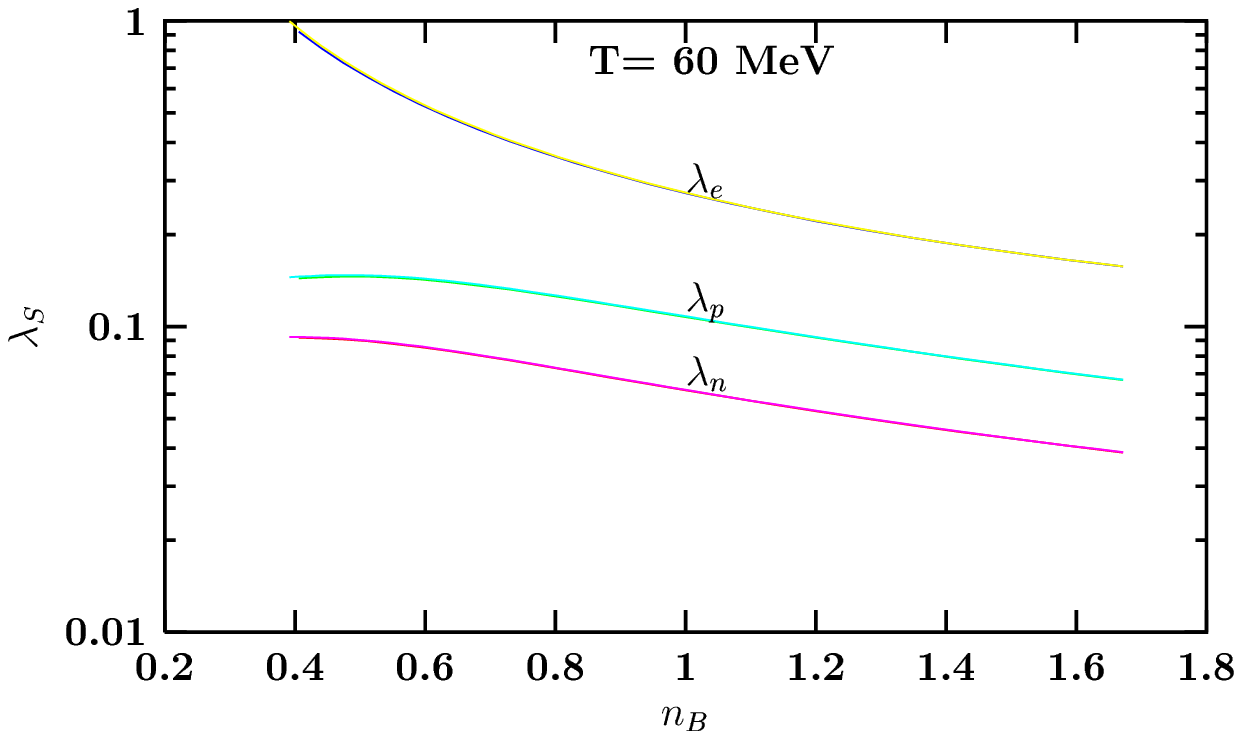}
	\hspace*{5 cm}
}

\hskip -8cm
\vskip -10cm
\caption{{\em Variation of neutrino mean free path
$\lambda_{S}$ (s=n, p, e) with $n_{B}$ for degenrate matter for 
T=5, 30 and 60 MeV and for $B=0$ and $5\times 10^4$. The left panel is for
untrapped matter $(Y_{\nu_e}=0)$ and the right panel is for trapped
matter $(Y_{L}=0.4)$
}}

\end{figure}

\pagebreak

\begin{figure}[ht]
\centerline{ 
\epsfxsize=20cm\epsfysize=40cm\epsfbox{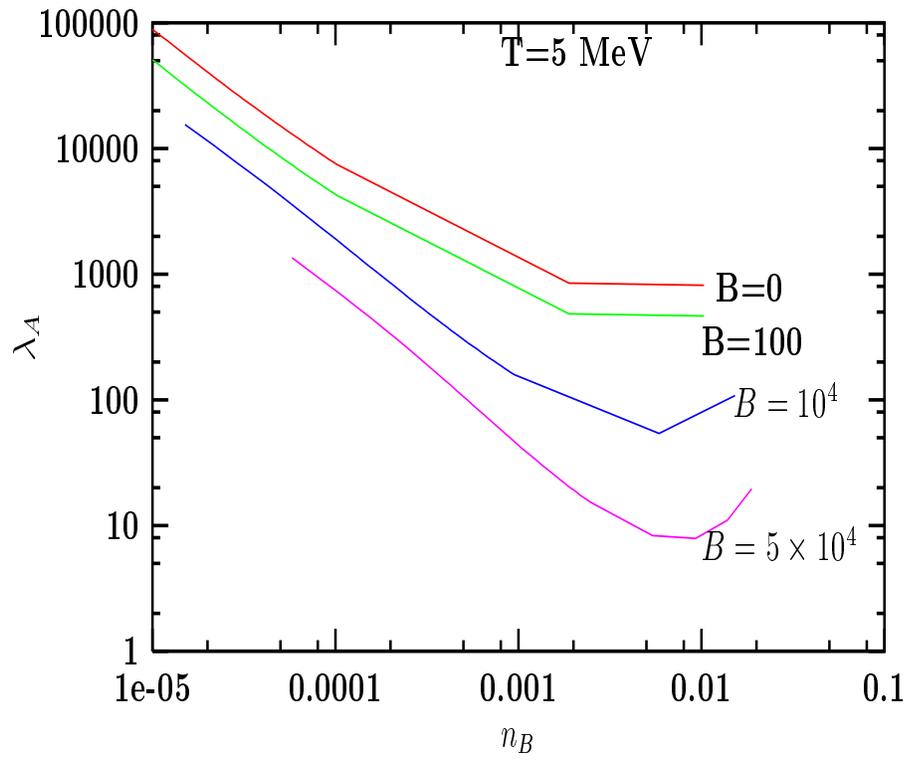}
}

\vskip -20cm
\caption{{\em Variation of $\lambda_{A}$ with $n_{B}$ for untrapped 
non-degenrate matter for T=5 MeV  and $B=0, 100, 10^4$ and $5\times 10^4$.
}}

\end{figure}

\pagebreak
\begin{figure}[ht]
\centerline{ 
\epsfxsize=10cm\epsfysize=26cm\epsfbox{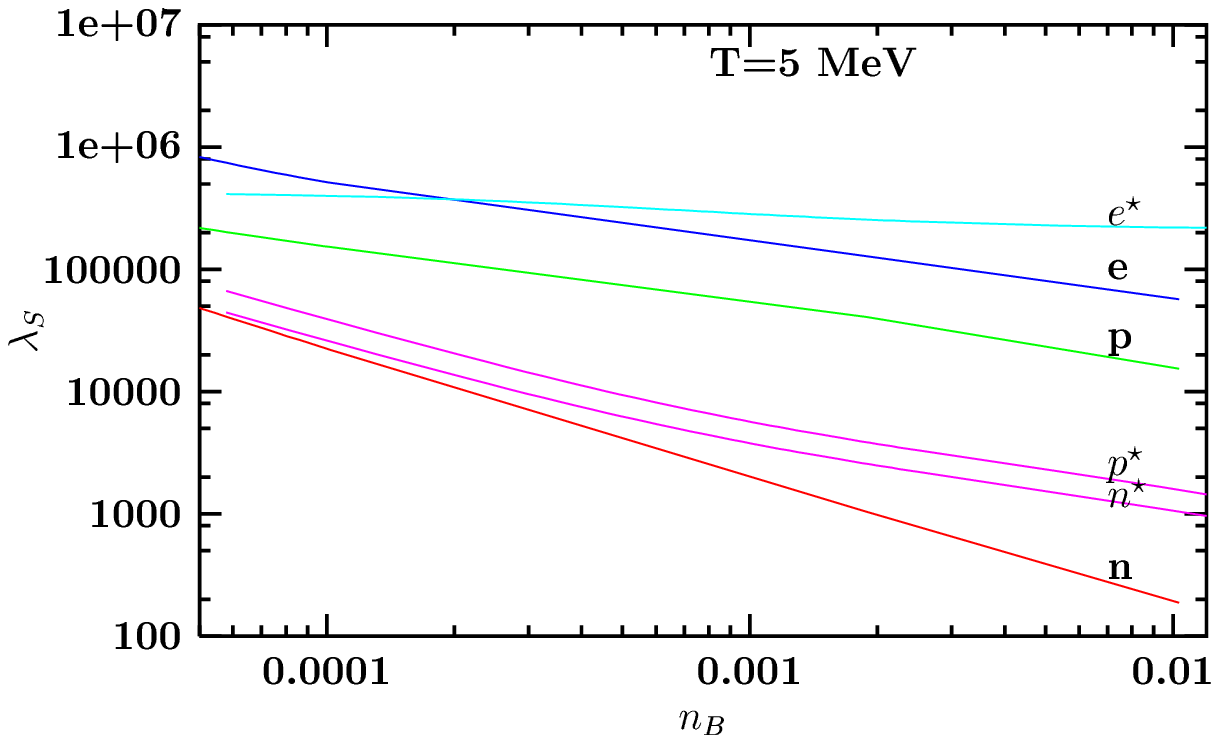}
	\hspace*{-4 cm}
\epsfxsize=10cm\epsfysize=26cm\epsfbox{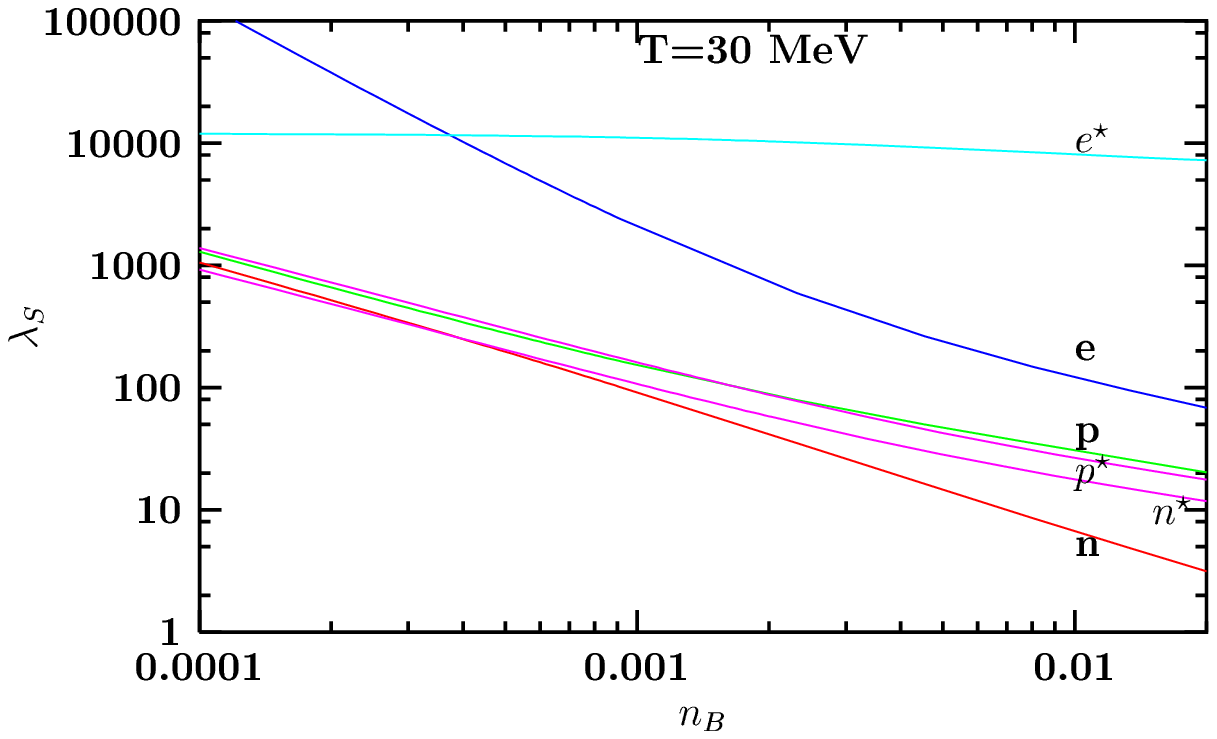}
}

\vskip -18.5cm
\centerline{ 
\epsfxsize=10cm\epsfysize=26cm\epsfbox{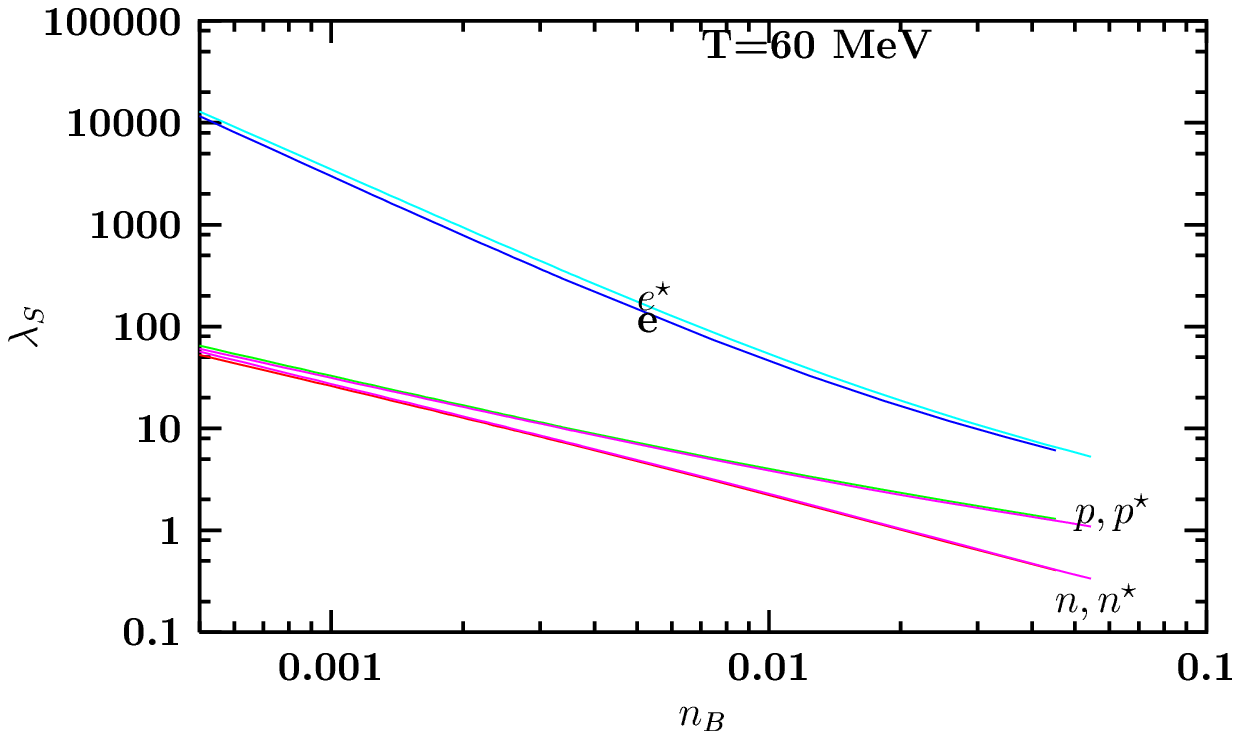}
}

\vskip -15cm
\caption{{\em Variation of neutrino scattering mean free path
$\lambda_{S}$ with $n_{B}$ for untrapped 
non-degenrate matter for T=5, 30 and 60 MeV.The curves 
labelled n, p, e  and  $n^\star$, $p^\star$,
$e^\star$
correspond to $\lambda_{n}$, $\lambda_{p}$ $\lambda_{e}$ 
for B=0 and $5\times 10^4$ respectively. 
}}

\end{figure}
\pagebreak


\end{document}